\documentclass[%
 reprint,
 amsmath,amssymb,
aps,
]{revtex4-2}

\usepackage{graphicx}% Include figure files
\usepackage{dcolumn}% Align table columns on decimal point
\usepackage{bm}% bold math
\usepackage[hidelinks]{hyperref}% add hypertext capabilities
\usepackage{siunitx}
\usepackage[version=4]{mhchem}
\usepackage{xcolor}
\usepackage[T1]{fontenc}
\usepackage{multirow}
\usepackage{booktabs}
\usepackage{url}

\usepackage{mathtools}
\DeclarePairedDelimiter\abs{\lvert}{\rvert}
\begin{document}

\preprint{APS/123-QED}

\title{Local indirect magnetoelectric coupling at twin walls in \ce{CaMnO3}}% Force line breaks with \\

\newcommand{\ntnu}{
    Department of Materials Science and Engineering, Faculty of Natural Sciences and Technology,
    NTNU Norwegian University of Science and Technology, NO-7491 Trondheim, Norway 
}

\author{Ida C. Skogvoll}
 \email{ida.c.skogvoll@ntnu.no}
  \affiliation{\ntnu}

\author{Benjamin A. D. Williamson}
  \affiliation{\ntnu}

\author{Sverre M. Selbach }%
 \email{selbach@ntnu.no}
 \affiliation{\ntnu}

\date{\today}% It is always \today, today,

\begin{abstract}
Ferroelastic twin walls in centrosymmetric perovskites can host emergent polar and magnetic properties forbidden in the bulk. We use density functional theory calculations to study the geometry and magnetic properties of ferroelastic domain walls in orthorhombic \ce{CaMnO3}, which belongs to the most common perovskite space group, $Pnma$. At the wall, the inherent inversion symmetry-breaking induces local polar distortions dependent on the wall geometry, which couple to the magnetic order through the octahedral distortions. Noncollinear calculations reveal enhanced out-of-plane magnetic moments on the \ce{Mn} atoms and a local, finite magnetization confined to the wall. Strain fields across twin walls thus give rise to coexistence of polarization and magnetization as well as magnetoelectric response that is absent and symmetry-forbidden in bulk \ce{CaMnO3}. We propose that magnetoelectric coupling and coexisting polarization and magnetization can emerge at twin walls in bulk centrosymmetric antiferromagnets.
\end{abstract}

%Old abstract: In this work, DFT+$U$ calculations were carried out on ferroelastic domain walls in the orthorhombic $Pnma$ perovskite \ce{CaMnO3} to investigate collinear and noncollinear magnetic properties and how they couple to strain-induced polar displacements. Ferroic domain walls display promising properties for nanoscale device applications, where local symmetry-breaking can give rise to phenomena not found in the bulk. In ferroelastic materials, twin walls exhibit inversion symmetry-breaking which can activate polar instabilities, yielding a net polarization. Noncollinear calculations show that specific domain wall symmetries enable a local magnetoelectric response and an enhanced out-of-plane net magnetization. The polarization profile at the wall is dependent on the spin configuration, as well as the local environment of octahedral distortions. The strain field present at ferroelastic domain walls can therefore alter the magnetic properties, making \ce{CaMnO3} a candiate material for realizing local multiferroicity in a single-phase material. As $Pnma$ is the most common space group for perovskites, isostructural twin walls in antiferromagnetic perovskites are suggested as a route to local multiferroicity and magnetoelectricity. Finally, local strain fields exist across all twin walls, inviting corresponding studies also of ferroelastic nonperovskites.

\maketitle

%\tableofcontents

\section{\label{sec:level1} Introduction }
Multiferroics are a class of materials that simultaneously exhibit at least two ferroic orders, and where the coexistence of magnetic order and ferroelectricity plays a particularly prominent role. The coupling of these orders is central to the development of multi-functional devices, where, for instance, the magnetization or polarization direction can be switched by an applied electric or magnetic field \cite{fiebig2016evolution}. However, the existence of ferroelectric and magnetic order parameters does not necessarily entail a strong coupling between the two. On the contrary, the common microscopic origins responsible for ferroelectric and magnetic orders are mutually exclusive by nature, and in the rare case that they do coexist, they are usually only weakly related \cite{Hill2000}.

Considerable efforts have been put into finding nondisruptive coupling mechanisms between the distortion mode responsible for polarization and the magnetic order. A notable contribution is improper ferroelectrics where the polar displacement results from complex spin-structures that break inversion symmetry \cite{Xiang2013}. Nonetheless, as frustrated magnetism commonly only features at very low temperatures and ferromagnetism is often achieved through double-exchange, materials that exhibit a coupling between polarization and ferromagnetism are exceedingly rare \cite{cheong2007multiferroics, aoyama2014giant, tokura2010multiferroics}. Other proposals consist of materials-by-design strategies for engineering weak ferromagnetism, where symmetry considerations are used to construct specific criteria for the interplay of spin and lattice degrees of freedom \cite{Ederer2005, Fennie2008, Fennie2011}. Both of these mechanisms are rooted in the Dzyaloshinskii-Moriya interaction (DMI), a relativistic correction to superexchange coupling, favoring noncollinear spin ordering \cite{dzyaloshinsky1958thermodynamic,Moriya1960}. 

A class of materials that lend themselves well to engineering multiferroics are the perovskite oxides. They have highly tunable properties deriving from the connection between structural, chemical, and spin degrees of freedom, as well as susceptibility of the \ce{ABO3} structure to compositional engineering \cite{Marthinsen2016, dagotto2008strongly, terakura2007magnetism, goodenough2014perspective}. Importantly, there is a prominent coupling between the characteristic rotations of the oxygen octahedra and the magnetic order in transition metal perovskites. Such rotational distortions in simple perovskites do not break inversion symmetry, and thus cannot contribute to a magnetoelectric coupling. However, in some layered perovskites a switchable magnetization by polarization reversal has been found, caused by a DMI and polar distortion emerging from the same octahedral distortion \cite{Benedek2011, Sun2021, Dey2025}. 

Because of their abundance, ferroelastic materials provide ample opportunities for investigating the relationship between magnetic order and octahedral rotations. Materials that exhibit either ferroelectricity or magnetism along with ferroelasticity are often not designated as being multiferroic, due to their less obvious technological potential. Hence, magnetism in ferroelastic materials has received less attention than in ferroelectrics. Nevertheless, in ferroelastic domain walls, inversion symmetry is inherently broken, and they exhibit an emergent strain field stabilizing a local metastable structure. As the order parameter switches from one domain to the other, the domain wall structure approaches a local symmetry that is distinctly different from the bulk, enabling novel functional properties that are not found within each domain \cite{meier2022ferroelectric}. 

\ce{CaMnO3} has the orthorhombic $Pnma$ symmetry in its ground state with a complex domain wall structure. This structure is characterized by the out-of-phase rotations of successive \ce{MnO6}-octahedra around two of the pseudocubic axes, and in-phase rotations around the remaining axis, corresponding to an $a^-a^-c^+$ Glazer tilt notation, where the three letters each represent an axis along with the tilting magnitude and phase ("$+$" being in-phase and "$-$" being out-of-phase). This leads to a $\sqrt{2}a_c\times\sqrt{2}a_c\times2a_c$ expansion of the unit cell with respect to the high-temperature cubic phase. Like many \ce{ABO3} structures it has a polar instability in the B-site cation, suppressed by the primary antiferrodistortive order parameter. This can be activated by straining the material, for example through anti-phase boundaries or twin walls \cite{barone2014improper,Kalinin2012, Gopalan2012, Stengel2017}.  Previous attempts to achieve ferroelectricity in the $Pnma$ phase have involved epitaxially straining thin films and building superlattices with mismatched layer constituents \cite{rabe2005theoretical, benedek2012polar, zhou2013strain}. Straining \ce{CaMnO3} can also induce other functionalities such as defect-formation control \cite{aschauer2014competition, aschauer2016interplay}. The space group of \ce{CaMnO3} allows canted antiferromagnetic order \cite{bousquet2011induced, rooj2025altermagnetism}, where the size of the canting is related to the octahedral rotations as well as the strength of the DMI and superexchange interactions \cite{bousquet2016non}. In other words, the bonding environment set by the octahedral rotation angles strongly affects the magnetic order through spin-phonon coupling \cite{Vanderbilt2012}. However, the magnetic order of \ce{CaMnO3} does not allow for a magnetoelectric response \cite{khalyavin2015antisymmetric}.

Here, we show that the intrinsic symmetry-breaking in two types of ferroelastic domain walls in \ce{CaMnO3} can facilitate a local net polarization and magnetization, and that these order parameters are coupled. The larger octahedral distortions at twin walls favor parallel spins compared to the bulk antiferromagnetism. Our results also reveal that the polarization profile and the amplitude of the net polar distortion are largely contingent on the specific magnetic configuration. Thus, the broken inversion symmetry indirectly lifts the suppression of the magnetoelectric effect, in the sense that the polarization couples to the magnetization through the primary order parameter. Noncollinear calculations show that a larger canting is induced at the domain wall, providing a net enhancement of the out-of-plane magnetization. 

The alteration of magnetic order at an interface has been treated in a few experimental and theoretical studies. However, these works involve superlattices, such as \ce{LaAlO3/PbTiO3} \cite{zhou2015ferroelectricity}, or multiferroics with a vanishing polarization at the domain wall \cite{Daraktchiev2010, geng2012collective, geirhos2020macroscopic}. In \ce{TbMnO3}, uncompensated spins at nonpolar twin walls lead to a net uncompensated in-plane magnetic moment \cite{farokhipoor2014artificial}. Strain-mediated coupling between magnetic and polar orders in ferroelastic domain walls is thus proposed as a way to achieve a magnetoelectric response in single-phase perovskite oxides. 

\section{\label{sec:methods} Computational methods}

DFT calculations were carried out using the \textsc{vasp} \cite{Kresse1996, Kresse1999} code and the PBEsol functional \cite{Perdew2008} with a spin-polarized GGA+$U$ implementation \cite{Dudarev1998}, to which a Hubbard $U$ correction of \SI{4}{eV} was applied to the \ce{Mn} $3d$ orbitals \cite{Andersen1991}. The projector augmented wave \cite{Blochl1994} method was used, where \ce{Ca}$(3s^2, 3p^6, 4s^2)$, \ce{Mn}$(3s^2,3p^6, 3d^6, 4s^1)$ and \ce{O}$(2s^2, 2p^4)$ were treated as valence electrons. A plane-wave cutoff energy of \SI{550}{eV} was applied, along with a $\varGamma$-centered $6\times 6\times 5$ k-point mesh for the primitive cell, with a corresponding k-point density employed for all other cell sizes. Atomic positions were relaxed until the residual forces were less than \SI{0.001}{eV/ \angstrom}. Domain wall structures with $8\times1\times1$ supercells of 40-atom unit cells were relaxed to below \SI{0.02}{eV/ \angstrom}. Phonon calculations were completed using the finite difference method implemented in the \textsc{phonopy} package, using $2\times2\times2$ and $3\times3\times3$ supercells for the 40-atom and primitive cells, respectively \cite{togo2023first, togo2023implementation}. The \textsc{isotropy} software suite was used for group-theoretical analysis \cite{isotropy}. Noncollinear magnetic calculations were conducted by including spin-orbit coupling self-consistently, both for determining the magnetocrystalline anisotropy energy and the magnetic wall configuration. \textsc{vesta} was used for visualization and illustration of the crystal structures \cite{Momma:db5098}.

\section{\label{sec:results} Results and discussion}
\begin{figure}[h!]
    \centering
    \includegraphics[width=1\linewidth]{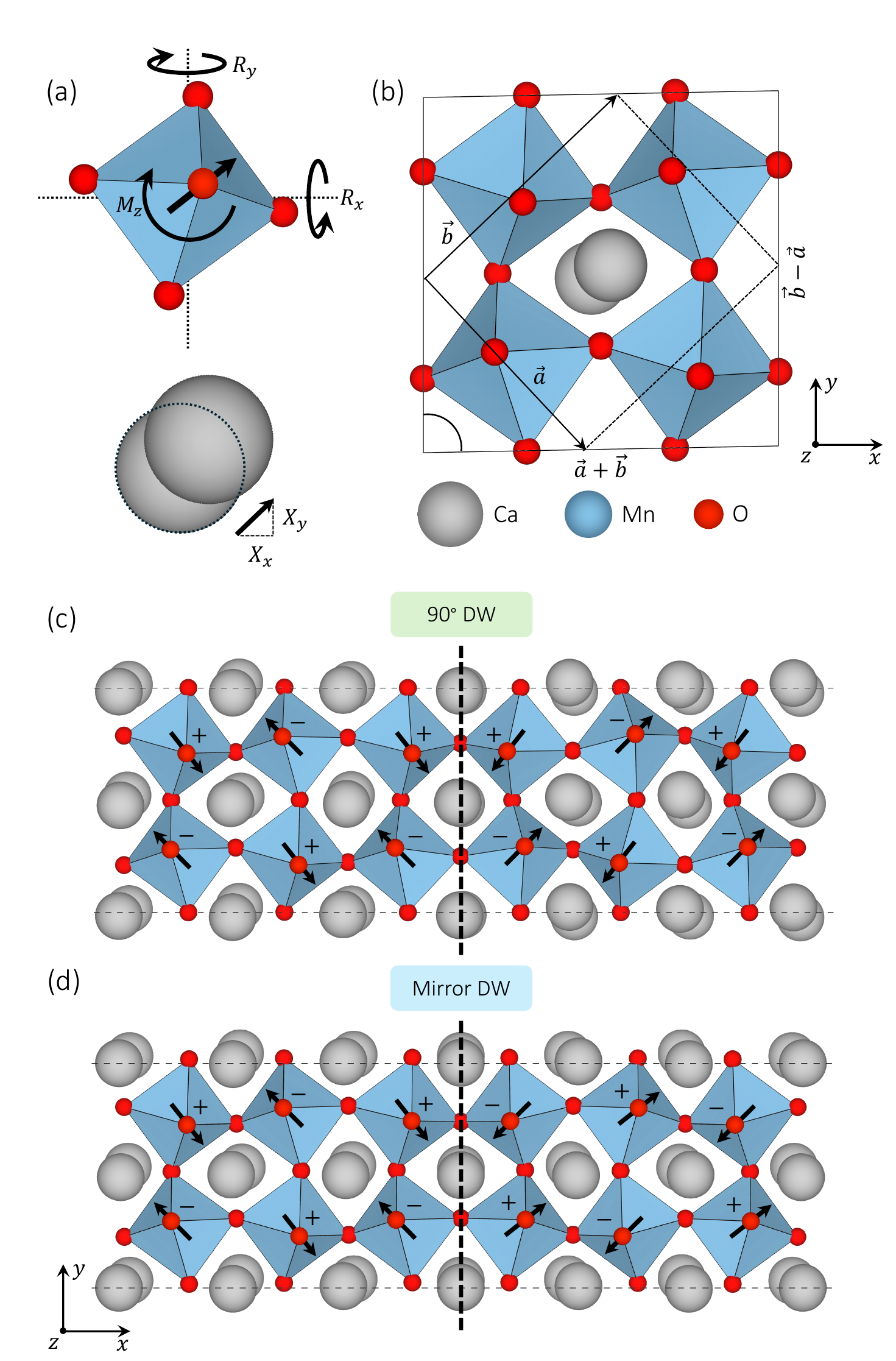}
    \caption{(a) The distortion modes present in the ground state of \ce{CaMnO3}. The top octahedron illustrates the rotational modes around each pseudocubic axis, namely the $R_4^+$ (here denoted $R_x$ and $R_y$ by their respective rotational axis), and $M_3^+$ (denoted $M_z$). The bottom figure displays the A-site displacement in the $xy$-plane, decoupled as an $X_y$ and $X_x$-mode along each axis. (b) Dashed lines mark the orthorhombic primitive cell, while solid lines display the strained 40-atom cell. The arc at the lower left vertex indicates the angle $90^\circ-\gamma$, where $\gamma$ is the twin angle. (c)-(d) Relaxed mirror and $90^\circ$ ferroelastic domain wall cells. Black arrows show the octahedral tilting direction resulting from the out-of-phase rotations, and "+" and "-" denote a clockwise or counter-clockwise in-phase rotation around the $z$-axis, prior to relaxation. The black dashed line illustrates the center position of the domain wall. Note that only half of the supercell is displayed here.}
    \label{fig:1}
\end{figure}
The optimized lattice constants $a$, $b$ and $c$ of the ground state primitive cell were determined to be $\SI{5.231}{\angstrom}$, $\SI{5.267}{\angstrom}$ and $\SI{7.410}{\angstrom}$, respectively. These values are lower than reported work using PBE, but close to experimental values of $\SI{5.271}{\angstrom}$, $\SI{5.280}{\angstrom}$ and $\SI{7.457}{\angstrom}$ \cite{terakura2007magnetism}. For the remainder of this work, the convention of defining the $c$-axis as the axis of in-phase rotations is used, corresponding to a $Pbnm$ setting. Coordinates of all Wyckoff positions are presented in Table \ref{tab:wyckoff}. The ground state magnetic ordering is denoted G-AFM, where spins are antiparallel between all neighboring \ce{Mn} atoms. The energies and geometric parameters associated with different collinear magnetic orders are presented in Table \ref{tab:magorders}, where it becomes apparent that in all cases parallel spins lead to a longer bond length and narrower bond angle, while antiparallel spins generate the opposite response. This is consistent with the parallel spin state favoring smaller orbital overlap, following the Goodenough-Kanamori-Anderson rules detailed in Section \ref{sec:results}(2). The C-AFM ordering, with parallel spins only in the $c$ direction is closest to the ground state, and the ferromagnetic (FM) is the least favorable. While these results highlight the notable spin-phonon coupling between the magnetic order and the octahedral rotations, there is only a few degrees separating a spin parallel versus an antiparallel configuration.

Octahedral tilting is the most common type of structural distortion in the perovskite phase, and can be defined in terms of the three-dimensional irreducible representations (irrep) $M_3^+$ and $R_4^+$ (using the convention where the B-site is at the origin) of the high-symmetry $Pm\bar{3}m$ space group, see Figure \ref{fig:1}(a). Investigating the phonon spectrum of the cubic phase, these distortions manifest themselves as imaginary modes at the $M$- and $R$-points of the Brillouin zone, along with a zone-center mode ($\Gamma_4^-$) representing the polar displacement of the B-site cation \cite{stokes2002group}. Applying the polar distortion to the cubic structure leads to no further energy-lowering. This is consistent with \ce{CaMnO3} having a B-site polar instability suppressed by the emergence of octahedral rotations at the antiferrodistortive phase transition, which are common competing structural instabilities in the perovskite family \cite{Klarbring2018, Rabe2009, aschauer2014competition, zhou2013strain}. Results from the phonon calculations are presented in Section I in the Supplemental material. As \ce{CaMnO3} has an orthorhombic crystal structure, it is ferroelastic by construction. The expanded pseudocubic unit cell, as shown in Figure \ref{fig:1}(b), exhibits a twin angle ($\gamma$) related to the primitive $a$ and $b$ lattice parameters. This can be calculated as $\gamma = 90 - 2\tan^{-1} (a/b)$, amounting to a twin angle of $0.4^\circ$. 

\begin{table}[b]
\centering
\caption{Wyckoff position and coordinates of all sites in the optimized ground state structure for \ce{CaMnO3}.}
\label{tab:wyckoff}
\begin{ruledtabular}
\begin{tabular}{lcccc}
Atom & Wyckoff position & x      & y      & z      \\ \colrule
Ca   & 4c               & 0.0076 & 0.0392 &    1/4    \\
Mn   & 4b               &     1/2 &   0    &     0    \\
O    & 4c               & 0.9290 & 0.4876 &    1/4    \\
     & 8d               & 0.2893 & 0.2884 & 0.0370  \\ 
\end{tabular}
\end{ruledtabular}
\end{table}

\begin{table*}
\centering
\caption{Energy and geometric parameters for \ce{CaMnO3} with different magnetic orderings in the primitive cell. The energy is calculated as the difference between the particular magnetic ordering and the G-AFM ground state. $\phi$ and  $\alpha$ denote the angle and average bond lengths in the Mn-O-Mn bonds of the $ab$-plane, while $\delta$ and $\beta$ represent all Mn-O-Mn angles and bonds in the $c$ direction (the axis of in-phase tilting).}
\label{tab:magorders}
\begin{ruledtabular}
\begin{tabular}{lcccccccc}
      & Energy (meV/u.c.) & \multicolumn{3}{c}{Lattice parameters (Å)}                            & \multicolumn{2}{c}{Mn-O-Mn bond angles ($^\circ$)} & \multicolumn{2}{c}{Bond lengths (Å)} \\
      & dE               & \multicolumn{1}{c}{a} & \multicolumn{1}{c}{b} & \multicolumn{1}{c}{c} & \multicolumn{1}{c}{$\phi$}        & $\delta$       & $\alpha$      & $\beta$     \\ \colrule
G-AFM & 0                & 5.231                  &           5.267       & 7.410                & 155.8                          & 157.0       & 1.898      & 1.891      \\
C-AFM & 18.07957         &    5.231              &         5.264         &   7.423               & 155.9                          & 156.2       & 1.898      & 1.896      \\
A-AFM & 34.85726         &   5.232               &             5.284    &    7.406               & 154.9                          & 157.1       & 1.905      & 1.889      \\
FM    & 90.91948         &   5.236               &           5.286       &   7.422               & 154.7                          & 155.8       & 1.907      & 1.898     
\end{tabular}
\end{ruledtabular}
\end{table*}

\subsubsection{Ferroelastic domain wall structure}

To establish the number of unique domain wall configurations, it is helpful to define the domain state in terms of the antipolar A-site distortion in the $\langle 110 \rangle_{\text{pc}}$ directions, see Figure \ref{fig:1}(a). This $X_5^+$-mode is an improper order parameter, coupled to the $M$ and $R$-modes through a trilinear coupling term, $\alpha Q_R Q_M Q_X$, in the free energy expansion. Its direction is therefore dictated by the local environment of octahedral rotations \cite{Zanolli2013}. Defining each rotational mode as $R_x (a^-a^0c^0)$, $R_y (a^0a^-c^0)$ and $M_z (a^0a^0c^+)$, the trilinear term can be decoupled as,
\begin{equation}
    F = \alpha_1 Q_{R_x}Q_{M_z}Q_{X_x} + \alpha_2 Q_{R_y}Q_{M_z}Q_{X_y},
\end{equation}
where $\alpha_1$ and $\alpha_2$ are arbitrary coupling coefficients \cite{oh2015experimental}. The shape of the unit cell is coupled to this A-site distortion such that modifying the ferroelastic domain state is associated with changing the sign of either $Q_{X_x}$ or $Q_{X_y}$. Thus, switching from one ferroelastic domain to another involves reversing the tilting phase of either a single $R$-mode or an $R$-mode and the $M$-mode in combination. As a result, there are four different domain wall configurations which can be categorized by whether the interfacing domains are related by a mirror plane or a $90^\circ$ rotation around the $z$-axis. Structures that lead to a disruption of the tilting pattern but no alteration of the strain direction result in structural anti-phase boundaries \cite{huang2016domain}. These will not be considered further. 

For this work, one domain wall structure resulting from a $90^\circ$ rotation and one exhibiting a mirror plane was constructed in the $(100)_{\text{pc}}$ plane. The former represents a switching of the $R_x$ and $M_z$ rotational modes, while the latter only switches the tilting phase of the first mode, covering the two kinds of tilting distortions expected at a ferroelastic wall. These walls will from here on be referred to as the "$90^\circ$ wall" and "mirror wall", see Figure \ref{fig:1}(c)-(d). The two walls in each supercell are structurally different interfaces although they have the same change in tilting phase. For each domain wall type, the domain wall formation energy was converged by testing dimensions $4\times 1\times 1$, $6\times 1\times 1$ and $8\times 1\times 1$, leading to values $E_{f, \text{DW}(90^\circ)} = \SI{163}{mJ/m^2}$ and $E_{f, \text{DW}(\text{mirror})} = \SI{67}{mJ/m^2}$. The former value closely agrees with previous calculated values for similar $90^\circ$ domain walls in \ce{CaTiO3} \cite{calleja2003trapping}. The lower formation energy of the latter domain wall is a natural consequence of less strain across a mirror plane. These formation energies are significantly larger than for the twin walls in \ce{SrTiO3} that are reported to be around $\SI{0.4}{mJ/m^2}$ \cite{Stengel2017}. However, the domain walls in \ce{SrTiO3} separate $a^0a^0c^-$ and $a^0a^0(-c^-)$ domains that only require minor structural rearrangement across the boundary and thus a far more subtle local strain field. In contrast, the domain wall energies in \ce{CaMnO3} are of similar magnitude to structural anti-phase boundaries in \ce{PbZrO3}, reported to be \SI{190}{mJ/m^2} \cite{wei2014ferroelectric}, and domain walls in prototypical proper ferroelectrics like \ce{LiNbO3} and \ce{PbTiO3}, with calculated energies of 120-$\SI{160}{mJ/m^2}$ \cite{zhang2023first} and $141-\SI{158}{mJ/m^2}$ \cite{eggestad2024mobile}. 
%are higher than expected values for proper ferroelectric domain walls in other perovskite species as these are associated with less structural deformation than twin walls (more citations needed).  
\begin{figure*}[ht]
    \centering
    \includegraphics[width=\textwidth]{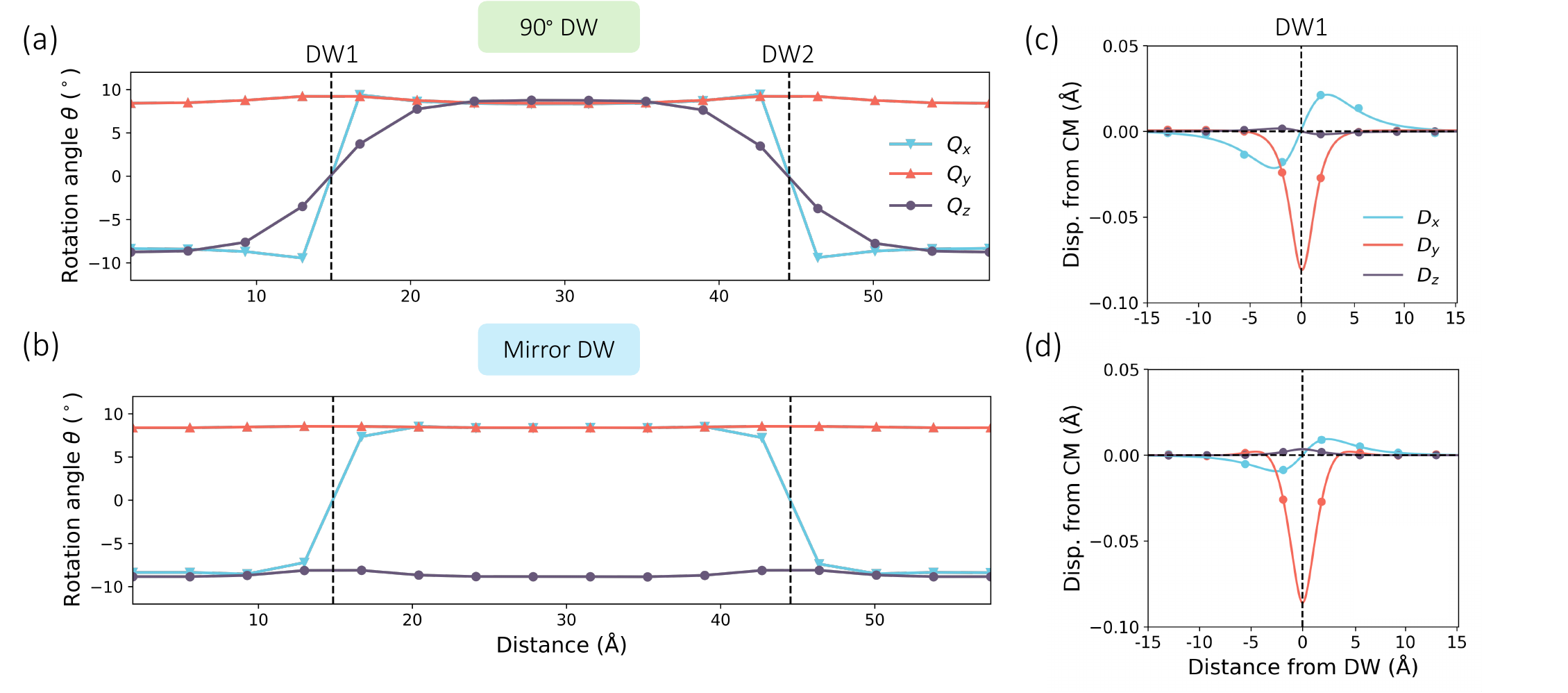}
    \caption{(a)-(b) Order parameter evolution calculated as the rotation angle around the $x$, $y$ and $z$-axes, as a function of the distance through the domain wall cell, for (a) a $90^\circ$ wall and (b) a mirror wall. The location of each domain wall is indicated by the dashed lines. (c)-(d) Polar distortion profiles of each domain wall type for the first domain wall of the cell, denoted "DW1". The distortion is calculated as the displacement of the \ce{Mn}-atom from the center of mass of the surrounding oxygen octahedron. DW2 shows the same trend, except that the direction of $D_y$ is reversed. For the mirror wall, $D_z$ is also reversed.}
    \label{fig:2}
\end{figure*}

Since the strain field across the domain wall stems from the phase change in the antiferrodistortive rotations of the oxygen octahedra, the primary order parameter $Q$ can be defined as the decomposed angle of rotation about each of the unit cell directions, coinciding with the pseudocubic axes and the chains of \ce{MnO6}-octahedra. For each octahedron, the rotation is explicitly defined in terms of the angle of four \ce{Mn-O} bonds in the corresponding plane normal to the rotation axis, relative to the cartesian coordinate system (or pseudocubic axes). The average angle around the $x$-axis is then defined as,
\begin{equation}
    \theta_x = \frac{1}{4}\sum_{\lambda=1}^4 \sin^{-1}( \mathbf{\hat{r}}_{yz,\ce{CM-O_\lambda}}\times \mathbf{\hat{r}'}_{yz,\ce{CM-O_\lambda}})
\end{equation}
where $\mathbf{\hat{r}}_{yz}$ is the unit vector pointing from the center of mass of the oxygen octahedra to each of the \ce{O} atoms within the $yz$-plane, indexed by $\lambda$, projected into the same plane. $\mathbf{\hat{r}'}_{yz}$ is the equivalent unit vector in the cubic cell, which coincides with the cartesian axes. If each position of an octahedron is indexed by the integers $(i,j,k)$, the decomposed order parameter for each layer parallel to the domain wall can be defined as,
\begin{align}
    Q_{R_x,R_y}(i) &= \frac{1}{4}\sum_{j,k} (-1)^{j+k}\theta_{x,y} \\
    Q_{M_z}(i) & = \frac{1}{4}\sum_{j} (-1)^{j}\theta_z
\end{align}
where the phase of octahedral rotations is accounted for. 

The resulting order parameter evolution throughout each supercell is displayed in Figure \ref{fig:2}(a)-(b). In both cases one of the order parameters pass through zero exhibiting a $\tanh(Q)$ relation, while the parameters that do not switch tilting phase follow a $\cosh(Q)$ profile, as expected from Landau theory \cite{salje1990phase}. The width of each domain wall was determined to be $w_{\text{DW}(90^\circ)} = \SI{19}{\angstrom}$ and $w_{\text{DW(mirror)}}= \SI{11}{\angstrom}$. These results also fit well with the larger strain field expected for the $90^\circ$ domain wall and the dissimilar formation energies. Narrow walls is in agreement with the inverse proportionality between the wall thickness and the critical temperature, $w \propto \abs{T-T_C}^{-1/2}$ \cite{salje1990phase}. Similar thin domain walls are also found in other ferroelastic oxides with the $Pnma$ structure \cite{zhao2019creating}.
\begin{figure}[h]
    \centering
    \includegraphics[width=1\linewidth]{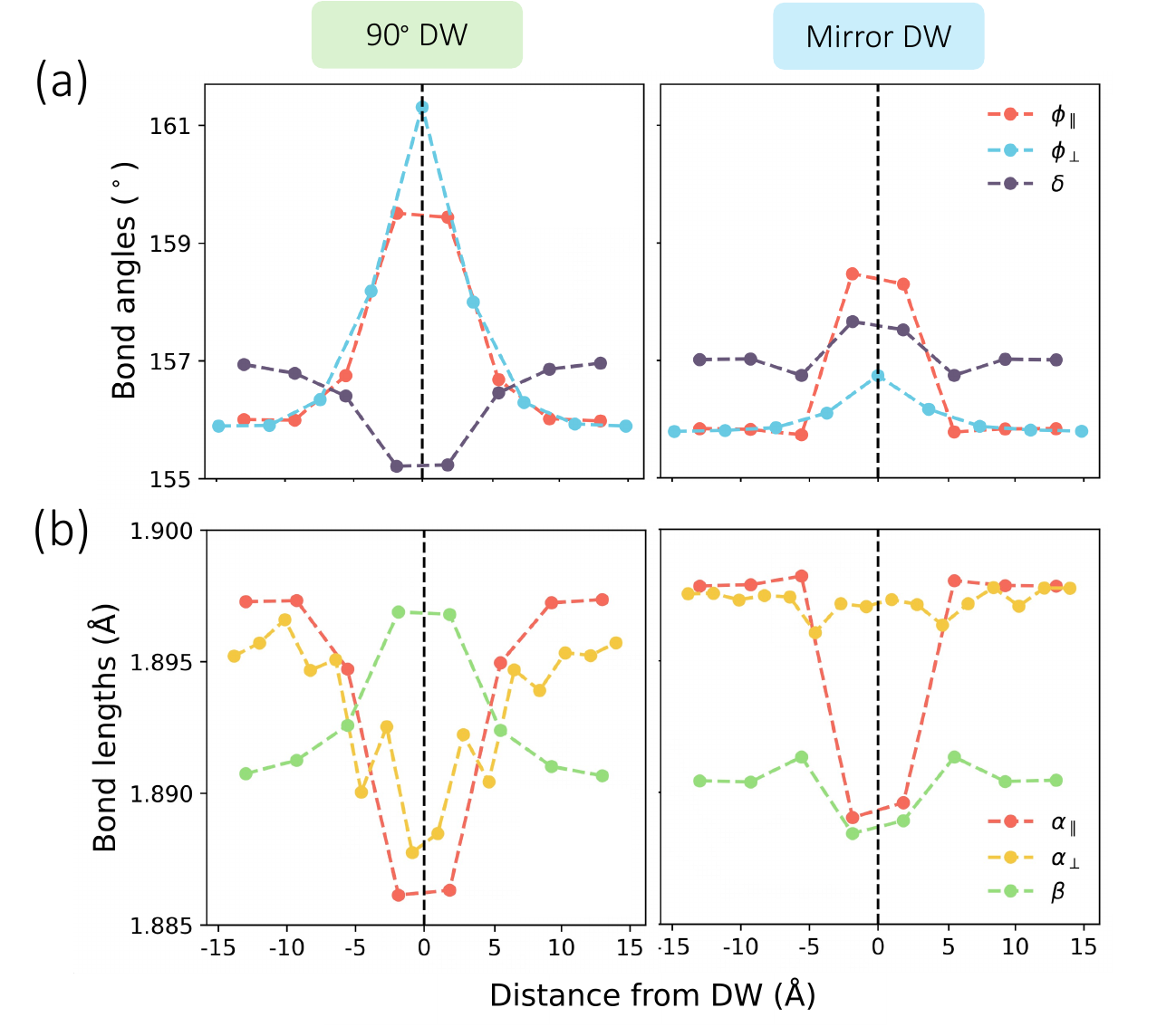}
    \caption{(a) Average Mn-O-Mn bond angles calculated as a function of distance from the first domain wall in each supercell. $\delta$ represents the angles around the axis of in-phase rotations ($z$), while $\phi_\|$ and $\phi_\perp$ are the in-plane angles, parallel and perpendicular to the domain wall. (b) Bond length alteration at each domain wall. $\alpha_\|$ and $\alpha_\perp$ denote the in-plane bond lengths parallel and perpendicular to the domain wall, while $\beta$ is the average bond angle in the $z$ direction. (a)-(b) All values converge to bulk values with the exception of $\alpha_{\perp,90^\circ}$ which is why these have been omitted from the plot.}
    \label{fig:3}
\end{figure}

The emergence of polar instabilities due to the vanishing of the primary order parameter is a well-known phenomenon and prevalent in ferroelastic domain walls \cite{barone2014improper, zhao2019creating}. The polarization originates from off-centering of the B-site ion. For \ce{CaMnO3} the polarization profile is therefore calculated as the displacement of the \ce{Mn}-atom from the center of mass of the encompassing oxygen octahedra, see Figure \ref{fig:2}(c)-(d). For a vanishing order parameter component, the ensuing polar distortion is of an antipolar character at each side of the domain wall, while for a nonvanishing and thus continuous tilting phase, a net polar distortion in one direction emerges. For the $90^\circ$ domain wall this results in a large antipolar distortion in the $x$ direction and a net \ce{Mn}-displacement in the $y$ direction. For the mirror domain wall, there is a similar profile for the $y$-component, while also displaying a small net displacement in the $z$ direction. The $D_y$ component of the polar distortion switches direction at each domain wall within each supercell, meaning that it is locked to the local octahedral environment, and thus of an improper nature. A detailed examination of the polarization profile in \ce{CaTiO3} and \ce{CaMnO3}, both in terms of A-site and B-site distortions is given in Ref. \cite{barone2014improper}. It is emphasized that the qualitative changes are different for each wall and that the magnitude of distortion is larger for the $90^\circ$ domain wall. 

\begin{table}[b]
\caption{Isotropy subgroups of $Pm\bar{3}m$ generated from the irreps $M_3^+$ and $R_4^+$ \cite{stokes2002group}. Only order parameter directions consistent with a well-defined Glazer notation is listed, i.e. where the order-parameter directions of the each irrep is not overlapping in terms of crystal axes.}
\label{tab:GT}
\begin{ruledtabular}
\begin{tabular}{lcccc}
\multirow{2}{*}{Irreps} & \multicolumn{2}{c}{Order param. dir.} & \multirow{2}{*}{Space group} & \multirow{2}{*}{Glazer notation} \\
                        & $M_3^+$               & $R_4^+$               &                              &                                  \\ \colrule
$M_3^+$                 & (a;0;0)               &                       & $P4/mbm$                     & $a^0a^0c^+$                    \\
                        & (a;a;0)               &                       & $I4/mmm$                     & $a^0b^+b^+$                    \\ 
                        & (a;a;a)               &                       & $Im\bar{3}$                     & $a^+a^+a^+$      \\ 
                        & (a;b;c)               &                       & $Immm$                     & $a^+b^+c^+$                     \\
$R_4^+$                 &                       & (a,0,0)               & $I4/mcm$                     & $a^0a^0c^-$                    \\
                        &                       & (a,a,0)               & $Imma$                       & $a^0b^-b^-$                    \\
                        &                       & (a,a,a)               & $R\bar{3}c$                       & $a^-a^-a^-$                    \\
                        &                       & (a,b,0)               & $C2/m$                       & $a^0b^-c^-$ \\
                        &                       & (a,b,b)               & $C2/c$                       & $a^-b^-b^-$  \\
                        &                       & (a,b,c)               & $P\bar{1}$                       & $a^-b^-c^-$                                                         \\
$M_3^+ \oplus R_4^+$    & (a;0;0)               & (0,0,b)               & $Cmcm$                       & $a^0b^-c^+$                    \\
                        & (a;0;0)                & (0,b,b)               & $Pnma$                       & $a^-a^-c^+$                    \\
                        & (a;0;0)               & (0,b,c)               & $P2_1/m$                     & $a^-b^-c^+$                    \\ 
                        & (a;a;0)               & (0,0,b)               & $P4_2/nmc$                     & $a^+a^+c^-$
\end{tabular}
\end{ruledtabular}
\end{table}
By determining the isotropy subgroups of each of the irreps $M_3^+$ and $R_4^+$, any possible evolution in the order parameter space is mapped out, thus listing potential switching pathways and intermediate domain wall symmetries, see Table \ref{tab:GT}. It is challenging to identify a specific intermediate symmetry from the domain wall structure itself because it is too narrow to encompass a full unit cell. There is therefore a distinction between the extrapolated space group and the actual symmetry of the unit cell centered at the wall. However, listing all possible intermediate symmetries that do not lead to over-lapping in- and out-of-phase rotations around a single axis, all resulting space groups have inversion symmetry and are nonpolar. The $90^\circ$ wall and mirror wall approximate a $Cmcm$ and $P4_2/nmc$ symmetry, respectively. For the former domain wall, the loss of in-phase rotations entails a complete re-alignment of the antipolar A-site distortion, while for the latter there is only a re-alignment in the direction normal to the wall. Additional distortions, in terms of A-site and B-site displacements must therefore be a purely flexoelectric effect. This is also in accordance with the larger polar distortion observed at the $90^\circ$ domain wall, which also exhibits the larger structural deformation. This effect of spontaneous polarization is predicted for \ce{CaTiO3} and other nonferroelectric perovskites to originate from the strain discontinuity at the boundary \cite{lu2025temperature}.

\subsubsection{Magnetic domain walls}
\label{sec:MWs}
\begin{figure*}[ht]
    \centering
    \includegraphics[width=\textwidth]{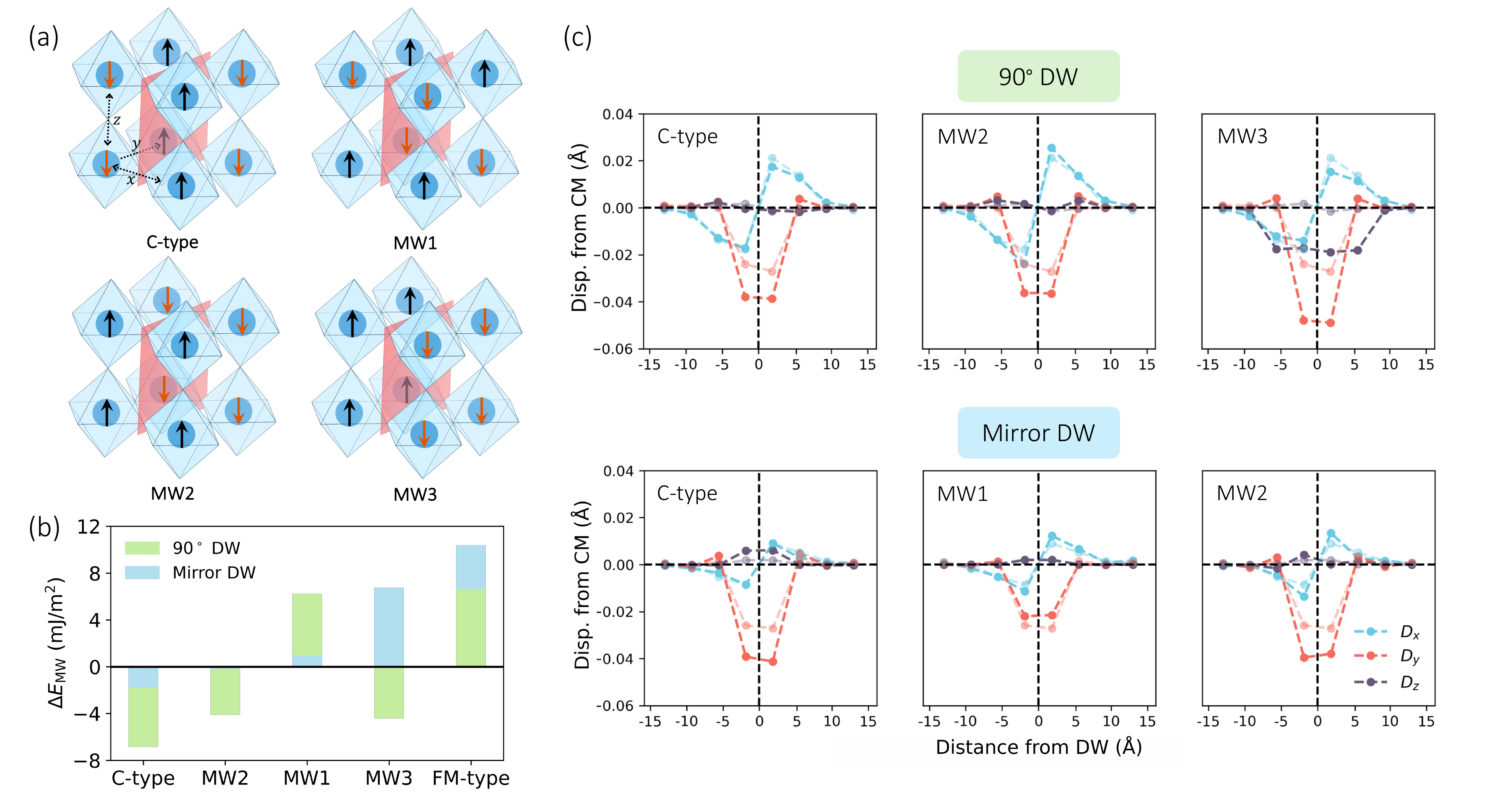}
    \caption{(a) Spin orders for the four lowest energy magnetic wall configurations. The cartesian axes coincide with those in Figure \ref{fig:1}. The position of the domain wall is indicated by the red plane. (b) Formation energies from placing a magnetic wall at the domain wall as opposed to a bulk structure, for a selection of spin orders, sorted with respect to the mirror domain wall. (c) Polar distortion profiles for the lowest energy magnetic walls for each respective domain wall type. For each magnetic wall the corresponding ground state distortion is shown at a higher transparency for comparison. The data is plotted without fits to hyperbolic functions to highlight the differences between the profiles.}
    \label{fig:4}
\end{figure*}
The magnetic order in \ce{CaMnO3} is described by the superexchange formalism, where the exchange interaction occurs through an intermediary nonmagnetic atom, in this case the \ce{O} atom connecting two \ce{Mn}. The exact strength and sign of the exchange interaction is based on the specific symmetry relations of the bonding orbitals, their electron occupancy and how strongly they overlap \cite{Anderson1950, KANAMORI195987, Anderson1959}. Spin ordering can thus be largely attributed to the \ce{Mn-O} bond lengths and angles, embodied in the local octahedral environment \cite{Terakura2010}. In terms of the Goodenough-Kanamori-Anderson rules, for \ce{d^3-d^3} cation configurations the exchange is most antiferromagnetic at $180^\circ$, increasingly more ferromagnetic for intermediate angles approaching $150^\circ$, and fully ferromagnetic at $90^\circ$ \cite{goodenough1963magnetism, lalena2020principles}. Octahedral distortions and deviating bond angles and lengths at ferroelastic domain walls can as a result stabilize a magnetic order different from the ground state found within bulk material. These are properties that are not captured through group-theoretical analysis or order-parameter profiles. The specific alterations of bond lengths and angles for each domain walls were therefore calculated as a function of the distance from the wall, shown in Figure \ref{fig:3}. It is evident that only the $90^\circ$ wall has a component of angle narrowing ($\delta$) and bond elongation ($\beta$) compatible with ferromagnetism, both in the $z$ direction. 

To test the susceptibility of the local octahedral environment towards different magnetic orders, a "magnetic wall" was imposed on top of the ferroelastic domain wall. Each collinear magnetic wall is then confined to the two octahedral layers adjacent to the center of each twin wall. Delineating each possible magnetic order, through parallel or antiparallel spins along each bonding direction ($x$, $y$ and $z$), amounts to eight different magnetic walls. Half of them correspond to a finite analogue of the regular bulk magnetic orders, and are therefore denoted for example "C-type" since they are not strictly three-dimensional, see Figure \ref{fig:4}(a). To evaluate the energetics of each magnetic wall, the formation energy was calculated according to the following relation, 
\begin{equation}
\begin{split}
    \Delta E_{\text{MW}} = \frac{1}{A_{DW}} \times & \big[(E_{\text{MW,DW}} - E_{\text{MW,bulk}})- \\
    & (E_{\text{G-AFM,DW}}-E_{\text{G-AFM,bulk}})\big],
\end{split}
\end{equation}
where the latter term corresponds to the formation energy of each ferroelastic domain wall, $E_{f,\text{DW}}$. This energy difference is a measure of the preference for a magnetic wall to reside within a twin wall compared to the bulk structure. It was ensured that the magnitude of the magnetic moments only showed small deviations from the bulk value of \qty{2.8}{\mu_B} for each \ce{Mn} atom. 

The resulting $\Delta E_{\text{MW}}$ for the four lowest energy magnetic walls along with the only magnetic wall of a net magnetic moment, FM-type, is presented in Figure \ref{fig:4}(b). For results concerning every wall type, see Section II(a) in the Supplemental material. In both the $90^\circ$ and mirror domain walls, it is found that the C-type magnetic order leads to the largest energy lowering, with $\Delta E_{\text{C-type},90^\circ} =$\SI{-6.85}{\milli\joule\per\meter\squared} and $\Delta E_{\text{C-type, mirror}} =$\SI{-1.78}{\milli\joule\per\meter\squared}. Accordingly, the local structure of both domain walls favors parallel spins in the out-of-plane direction and opposes parallel spins in the remaining $xy$-plane. The second and third lowest formation energies result from the MW2 and MW3 configuratiosn, which both also have parallel spins in the $z$ direction, while the highest energy configurations are the A-type and FM magnetic walls. 

From these results, the cost of parallel versus antiparallel spins in each direction can be inferred, seeing that the $z$ direction parallel spins lead to the largest energy lowering, while a larger degree of spin alignment in the $x$ and $y$ directions leads to an energy gain. This can be understood in terms of the geometric considerations stated in the Goodenough-Kanamori formalism. As previously discussed, it is clear that for the $90^\circ$ domain wall, the bond angles become smaller in the $z$ direction while straightening in the $x$ and $y$ directions, see Figure \ref{fig:3}. For the mirror domain wall, all angles become larger but with a smaller magnitude. In terms of bond lengths, an inverted but similar trend is found. This correlates well with the more significant energy reduction for the C-type magnetic wall and with the larger overall variance in formation energies, seen for the $90^\circ$ domain wall. Although none of these magnetic walls give a lower energy than a fully G-AFM ground state, the energetic trends still demonstrate the propensity of the domain wall geometry towards a different magnetic ordering, implying a greatly locally enhanced magnetic susceptibility. For all the magnetic wall orderings, the same quantitative alterations in angle and bond deformation as seen in the bulk, is observed at the domain walls. 

Decomposing the magnetic order parameter in terms of the irreps of the $Pnma$ space group, it is found that there are four sets of symmetry allowed magnetic orders that also allow canting. However none of these allow a magnetoelectric coupling because inversion symmetry is still conserved. This is also consistent with there being no third order coupling allowed between the magnetic order and the octahedral tilting modes for B-site magnetism. The only way to activate a magnetoelectric response is therefore through inversion symmetry-breaking \cite{bousquet2011induced, khalyavin2015antisymmetric, senn2018group}. Calculating the displacement of \ce{Mn} from the center of mass of each oxygen octahedra, we find that there is such an induced effect also in the twin walls of \ce{CaMnO3}, see Figure \ref{fig:4}(c). 

For almost every magnetic wall there is an enhancement of the polar distortion with the degree of spin alignment. Although the distortion profiles are qualitatively different between the two ferroelastic domain wall types, for every magnetic wall with parallel spins in the $z$ direction, an enhancement of the net $D_y$ polar distortion is observed both in the $90^\circ$ and mirror domain wall. Particularly, for the "MW3" wall, the polar distortion is almost twice as large. Some spin orders also break the antipolar or polar order of the displacements. For polar distortion profiles for each magnetic order, see Section IIb in the Supplemental material.  

Although it is difficult to disentangle the exact coupling between these order parameters due to the nature of the symmetry-breaking at the domain wall, it is evident that the emergence of a polar distortion lifts the restriction forbidding a magnetoelectric coupling. This is however not a direct coupling, but an indirect link through the primary order parameter. 

\subsubsection{Noncollinear magnetic calculations}
\begin{figure}[h]
    \centering
    \includegraphics[width=1\linewidth]{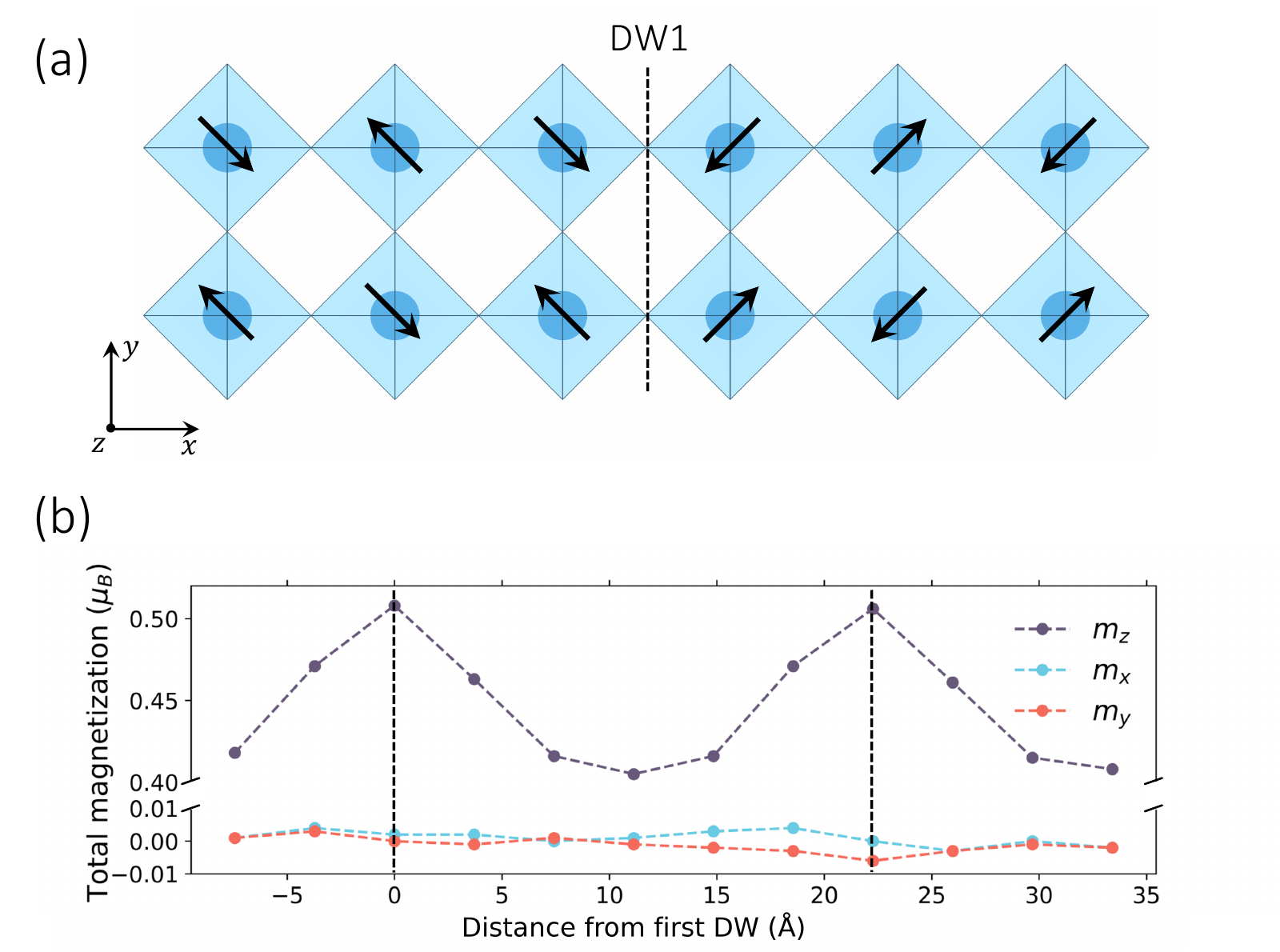}
    \caption{(a) The initial magnetic order prior to the self-consistent spin rotation. The magnetic domain wall corresponds to a "mirror" symmetric interface (if the spins are not treated as axial vectors). (b) The resulting average magnetization of one unit cell consisting of eight \ce{Mn} atoms. While $m_x$ and $m_y$ cancel each other, there is a net magnetization, $m_z$, in the out-of-plane direction. Note the break in the $y$-axis scale.}
    \label{fig:5}
\end{figure}
In the preceding calculations, all spins were restricted to a collinear setting with no coupling to the lattice degrees of freedom. From previous work on \ce{CaMnO3} it was shown that the ground state is a canted G-AFM order with a weak ferromagnetic component, specifically the $G_x A_y F_z$ spin configuration \cite{bousquet2011induced}. This means that the net magnetization is in the direction found to exhibit the highest susceptibility towards a parallel spin configuration in a ferroelastic domain wall, as well as a notable coupling to the net polar displacement. It is therefore reasonable to anticipate a coupling between the canted antiferromagnetic ground state and the local symmetry at the twin wall, driving a larger degree of spin alignment such as observed in the collinear calculations. These effects are expected to be more prominent for the $90^\circ$ domain wall as it exhibits the largest and most favorable distortions, which is why the analysis is limited to this wall type. 

To determine the most stable spin orientation of a G-AFM ordering, bulk calculations were conducted confirming the correct symmetry-allowed canting pattern for each crystallographic direction and that the ground state is $G_x A_y F_z$. Assuming that the easy-axis is conserved within each domain, this magnetic ordering was imposed on a $6\times 1 \times 1$ supercell with the relaxed geometry, so that there is a $90^\circ$ rotation of the spin direction at the domain wall. This allows two options where the spins at the interface are either mirrored or rotated. The mirrored configuration was found to give the lowest energy by \SI{0.01}{meV/{f.u.}}, and is illustrated in Figure \ref{fig:5}(a). 

By optimizing spin directions while keeping atomic positions fixed, we observed an increase in the canting angle both in the $xy$-plane from $1.5^\circ$ to $2.3^\circ$, while in the $z$ direction the canting angle only increased from $1.0^\circ$ to $1.3^\circ$. See Section III in the Supplemental material for an outline of the domain wall canting in terms of spherical coordinates $(\phi, \theta)$. As a result, there is a net increase in the magnetization in the out-of-plane direction, while the in-plane magnetization still cancels to zero, see Figure \ref{fig:5}(b). This applies to both domain walls despite the reversal of the net polar distortion at each respective wall. The energy of a single magnetic domain is also lower in energy than the two separate domains, probably because the cost of the magnetic interface is higher than the magnetic anisotropy energy for a supercell. 

Detection of the enhanced out-of-plane magnetization requires experimental techniques that can spatially resolve magnetic moments at domain wall dimensions below the Néel temperature ($T_N$). This could be accomplished through direct imaging techniques such as spin-polarized scanning tunneling microscopy (SP-STM) \cite{cheong2020seeing}, magnetic force microscopy (MFM) \cite{geng2014direct} or scanning nitrogen-vacancy (NV) magnetometry \cite{chauleau2020electric}. Lorentz transmission electron microscopy (L-TEM) could also be able to resolve ferromagnetic moments at a spatial resolution smaller than the width of a twin wall \cite{han2019topological}.
%, and magnetoelectric force microscopy (MeFM) has previously been used to detect magnetoelectric response \cite{geng2014magnetoelectric}.

The spin canting is a result of a complex interaction between the magnetocrystalline anisotropy energy (single-ion anisotropy), the superexchange coupling and DMI, which all can depend on the displacement of oxygen through octahedral rotations \cite{Zhou2020, Dagotto2006}. Such analysis becomes increasingly difficult with lower structural symmetry. However, it has been found that the amplitude of the canting generally depends on the ratio of the DMI and superexchange interaction, where a larger octahedral distortion involves a smaller $J$ and a larger canting \cite{Weingart2012}. The $G_x A_y F_z$ magnetic order parameter can also be decomposed in bi-linear free energy terms, resulting in relations $M_3^+ \propto G_x A_y$ and $R_4^+ \propto G_x F_z$ \cite{khalyavin2015antisymmetric}. Both of these considerations imply that there is a larger canting associated with an increase in the $Q_x$ and $Q_y$ order parameters, as observed at the $90^\circ$ domain wall.  This is also in accordance with the largest $D_y$ enhancement resulting from parallel spins in the out-of-plane direction, since these rotations have the strongest effect for in-plane rotation angles. Repeating these single-point calculations with an incrementally increased displacement of \ce{Mn} did not lead to any larger spin canting at the wall, proving that the polar response to the magnetic order is contingent on the rotational degrees of freedom. 

Thus, the inversion symmetry-breaking at the domain wall allows for a magnetoelectric effect. However, this is a strain-mediated response where the magnetic order couples to the polarization through a magnetoelastic-flexoelectric relation. Such a coupling response through an octahedral distortion is also observed for example in bulk Ruddlesden-Popper \ce{Ca3Mn2O7} \cite{Fennie2011, zemp2024magnetoelectric}. \ce{CaMnO3} offers an untraditional route towards this effect. In addition, the abundance of ferroelastic nonpolar antiferromagnets, e.g. rare earth chromites, manganites and ferrites among just perovskites, suggest the existence of more ideal ferroelastic domain walls with respect to local ferromagnetism, improper polarization and magnetoelectric coupling.

% An enhanced out-of-plane magnetization associated with a polar domain wall carries potential for nanoelectronics, e.g. by electric field control of the magnetic state. Detection of enhanced out-of-plane magnetization requires experimental techniques that can spatially resolve magnetic moments at domain wall dimensions below the Néel temperature ($T_N$). This could be accomplished through direct imaging techniques such as spin-polarized scanning tunneling microscopy (SP-STM) \cite{cheong2020seeing}, magnetic force microscopy (MFM)\cite{geng2014direct} or scanning nitrogen-vacancy (NV) magnetometry \cite{chauleau2020electric}. Lorentz transmission electron microscopy (L-TEM) could also be able to resolve ferromagnetic moments at a spatial resolution smaller than the width of a twin wall \cite{han2019topological}, and magnetoelectric force microscopy (MeFM) has previously been used to detect magnetoelectric response \cite{geng2014magnetoelectric}. The magnetoelectric coupling in the domain wall could also potentially be observed in macroscopic measurements of the dielectric response as a function of temperature. 

\section{Conclusion}
Ferroelastic twin walls in orthorhombic \ce{CaMnO3} are shown to host local polarization and magnetization couplings that are symmetry-forbidden in the bulk. Broken inversion symmetry from gradients in octahedral tilting induce domain wall-centered polarization, while the resulting DMI vectors enhance out-of-plane \ce{Mn} magnetization that is confined to the wall. This coupling between the local polarization and weak ferromagnetism strongly support a strain-mediated magnetoelectric response at ferroelastic twin walls through the octahedral distortions. It is dependent on the domain wall geometry, and strongest for twin walls associated with the largest strain gradient. 

These results should also be general for all $Pnma$ structures, which is the most common space group among the the broad family of perovskites. In addition, we propose that ferroelastic twin walls in other bulk centrosymmetric antiferromagnets can induce local polarization and magnetization. Our results suggest that large strain gradients at domain walls give the strongest effect, implying that good candidates have narrow domain walls and large distortions relative to high-symmetry aristotype phases. Such features are typically found in compounds with high ferroelastic Curie temperatures making e.g. \ce{LaFeO3} a prime candidate. 

%old conclusion: First-principles calculations show that ferroelastic domain walls in \ce{CaMnO3} display local symmetry-breaking that activates magnetoelectric response forbidden by the bulk symmetry. This yields an enhanced out-of-plane ferromagnetic moment, which is coupled to the spontaneous polarization. This effect is dependent on the domain wall geometry, and strongest for twin walls associated with the largest strain gradient. These calculations strongly support the concept of magnetoelectric coupling at twin walls in magnetic and nonferroelectric perovskites with tilted octahedra. 
%
%The domain wall geometries used in this work are derived generally for $Pnma$-structures, and should be translatable to other materials systems that may be more favorable in terms of magnetic order or octahedral distortions. This could for example include twin walls in \ce{LaFeO3}, where the superexchange coupling is based on the stronger $\pi$-bonding between the \ce{Fe}-orbitals, yielding a higher $T_N$. A stronger effect might also be achieved through compositional or strain engineering of \ce{CaMnO3}, particularly if this permits a larger tilting amplitude. 

\begin{acknowledgments}
\vspace{-0.5mm}
The authors thank Nicola Spaldin, Manfred Fiebig, Jan Schulthei{\ss}, Sophie F. Weber and Kristoffer Eggestad for valuable insight and helpful discussions. Financial support for this work was received from the Research Council of Norway through project 302506. Computational resources were provided by UNINETT Sigma2 through the project NN9264K. 

\end{acknowledgments}
\bibliography{apssamp}% Produces the bibliography via BibTeX.

@article{cheong2007multiferroics,
  title={Multiferroics: a magnetic twist for ferroelectricity},
  author={Cheong, Sang-Wook and Mostovoy, Maxim},
  journal={Nat. Mater.},
  volume={6},
  number={1},
  pages={13--20},
  year={2007},
  publisher={Nature Publishing Group UK London},
  doi={https://doi.org/10.1038/nmat1804}
}

@article{Fennie2008,
  title = {Ferroelectrically {I}nduced {W}eak {F}erromagnetism by {D}esign},
  author = {Fennie, Craig J.},
  journal = {Phys. Rev. Lett.},
  volume = {100},
  issue = {16},
  pages = {167203},
  numpages = {4},
  year = {2008},
  month = {Apr},
  publisher = {American Physical Society},
  doi = {10.1103/PhysRevLett.100.167203},
  url = {https://link.aps.org/doi/10.1103/PhysRevLett.100.167203}
}

@article{Hill2000,
author = {Hill, Nicola A.},
title = {Why {A}re {T}here so {F}ew {M}agnetic {F}erroelectrics?},
journal = {J. Phys. Chem. B},
volume = {104},
number = {29},
pages = {6694-6709},
year = {2000},
doi = {10.1021/jp000114x}
}

@article{fiebig2016evolution,
  title={The evolution of multiferroics},
  author={Fiebig, Manfred and Lottermoser, Thomas and Meier, Dennis and Trassin, Morgan},
  journal={Nat. Rev. Mater},
  volume={1},
  number={8},
  year={2016},
  publisher={Nature Publishing Group},
  pages={16046},
  doi ={https://doi.org/10.1038/natrevmats.2016.46}}

@article{Ederer2005,
  title = {Weak ferromagnetism and magnetoelectric coupling in bismuth ferrite},
  author = {Ederer, Claude and Spaldin, Nicola A.},
  journal = {Phys. Rev. B},
  volume = {71},
  issue = {6},
  pages = {060401},
  numpages = {4},
  year = {2005},
  month = {Feb},
  publisher = {American Physical Society},
  doi = {10.1103/PhysRevB.71.060401},
  url = {https://link.aps.org/doi/10.1103/PhysRevB.71.060401}
}

@article{Sun2021,
  title = {Hybrid improper ferroelectricity and magnetoelectric coupling in a two-dimensional perovskite oxide},
  author = {Zhou, Ying and Chen, Zefeng and Wu, Zongshuo and Shen, Xiaofan and Wang, Jianli and Zhang, Junting and Sun, Hui},
  journal = {Phys. Rev. B},
  volume = {103},
  issue = {22},
  pages = {224409},
  numpages = {7},
  year = {2021},
  month = {Jun},
  publisher = {American Physical Society},
  doi = {10.1103/PhysRevB.103.224409},
  url = {https://link.aps.org/doi/10.1103/PhysRevB.103.224409}
}

@article{Benedek2011,
  title = {Hybrid {I}mproper {F}erroelectricity: {A} {M}echanism for {C}ontrollable {P}olarization-{M}agnetization {C}oupling},
  author = {Benedek, Nicole A. and Fennie, Craig J.},
  journal = {Phys. Rev. Lett.},
  volume = {106},
  issue = {10},
  pages = {107204},
  numpages = {4},
  year = {2011},
  month = {Mar},
  publisher = {American Physical Society},
  doi = {10.1103/PhysRevLett.106.107204},
  url = {https://link.aps.org/doi/10.1103/PhysRevLett.106.107204}
}

@article{Kresse1996,
  title = {Efficient iterative schemes for \textit{ab initio} total-energy calculations using a plane-wave basis set},
  author = {Kresse, G. and Furthm\"uller, J.},
  journal = {Phys. Rev. B},
  volume = {54},
  issue = {16},
  pages = {11169--11186},
  numpages = {0},
  year = {1996},
  month = {Oct},
  publisher = {American Physical Society},
  doi = {10.1103/PhysRevB.54.11169},
  url = {https://link.aps.org/doi/10.1103/PhysRevB.54.11169}
}

@article{Kresse1999,
  title = {From ultrasoft pseudopotentials to the projector augmented-wave method},
  author = {Kresse, G. and Joubert, D.},
  journal = {Phys. Rev. B},
  volume = {59},
  issue = {3},
  pages = {1758--1775},
  numpages = {0},
  year = {1999},
  month = {Jan},
  publisher = {American Physical Society},
  doi = {10.1103/PhysRevB.59.1758},
  url = {https://link.aps.org/doi/10.1103/PhysRevB.59.1758}
}

@article{Perdew2008,
  title = {Restoring the {D}ensity-{G}radient {E}xpansion for {E}xchange in {S}olids and {S}urfaces},
  author = {Perdew, John P. and Ruzsinszky, Adrienn and Csonka, G\'abor I. and Vydrov, Oleg A. and Scuseria, Gustavo E. and Constantin, Lucian A. and Zhou, Xiaolan and Burke, Kieron},
  journal = {Phys. Rev. Lett.},
  volume = {100},
  issue = {13},
  pages = {136406},
  numpages = {4},
  year = {2008},
  month = {Apr},
  publisher = {American Physical Society},
  doi = {10.1103/PhysRevLett.100.136406},
  url = {https://link.aps.org/doi/10.1103/PhysRevLett.100.136406}
}

@article{Dudarev1998,
  title = {Electron-energy-loss spectra and the structural stability of nickel oxide:  An {LSDA}+{U} study},
  author = {Dudarev, S. L. and Botton, G. A. and Savrasov, S. Y. and Humphreys, C. J. and Sutton, A. P.},
  journal = {Phys. Rev. B},
  volume = {57},
  issue = {3},
  pages = {1505--1509},
  numpages = {0},
  year = {1998},
  month = {Jan},
  publisher = {American Physical Society},
  doi = {10.1103/PhysRevB.57.1505},
  url = {https://link.aps.org/doi/10.1103/PhysRevB.57.1505}
}

@article{Blochl1994,
  title = {Projector augmented-wave method},
  author = {Bl\"ochl, P. E.},
  journal = {Phys. Rev. B},
  volume = {50},
  issue = {24},
  pages = {17953--17979},
  numpages = {0},
  year = {1994},
  month = {Dec},
  publisher = {American Physical Society},
  doi = {10.1103/PhysRevB.50.17953},
  url = {https://link.aps.org/doi/10.1103/PhysRevB.50.17953}
}

@article{Zanolli2013,
  title = {Electric control of the magnetization in \ce{BiFeO3}/\ce{LaFeO3} superlattices},
  author = {Zanolli, Zeila and Wojde\l{}, Jacek C. and \'I\~niguez, Jorge and Ghosez, Philippe},
  journal = {Phys. Rev. B},
  volume = {88},
  issue = {6},
  pages = {060102},
  numpages = {5},
  year = {2013},
  month = {Aug},
  publisher = {American Physical Society},
  doi = {10.1103/PhysRevB.88.060102},
  url = {https://link.aps.org/doi/10.1103/PhysRevB.88.060102}
}

@article{huang2016domain,
  title={Domain topology and domain switching kinetics in a hybrid improper ferroelectric},
  author={Huang, F-T and Xue, F and Gao, B and Wang, LH and Luo, X and Cai, W and Lu, X-Z and Rondinelli, JM and Chen, LQ and Cheong, S-W},
  journal={Nat. Commun.},
  volume={7},
  number={1},
  pages={11602},
  year={2016},
  publisher={Nature Publishing Group UK London},
  doi={https://doi.org/10.1038/ncomms11602}
}

@article{oh2015experimental,
  title={Experimental demonstration of hybrid improper ferroelectricity and the presence of abundant charged walls in \ce{(Ca,Sr)3Ti2O7} crystals},
  author={Oh, Yoon Seok and Luo, Xuan and Huang, Fei-Ting and Wang, Yazhong and Cheong, Sang-Wook},
  journal={Nat. Mater.},
  volume={14},
  number={4},
  pages={407--413},
  year={2015},
  publisher={Nature Publishing Group UK London},
  doi={https://doi.org/10.1038/nmat4168}
}

@article{stokes2002group,
  title={Group-theoretical analysis of octahedral tilting in ferroelectric perovskites},
  author={Stokes, Harold T and Kisi, Erich H and Hatch, Dorian M and Howard, Christopher J},
  journal={Acta Crystallogr. Sect. B},
  volume={58},
  number={6},
  pages={934--938},
  year={2002},
  publisher={International Union of Crystallography},
  doi={https://doi.org/10.1107/S0108768198004200}
}

@article{togo2023implementation,
  title={Implementation strategies in phonopy and phono3py},
  author={Togo, Atsushi and Chaput, Laurent and Tadano, Terumasa and Tanaka, Isao},
  journal={J. Phys.: Condens. Matter},
  volume={35},
  number={35},
  pages={353001},
  year={2023},
  publisher={IOP Publishing},
  doi={10.1088/1361-648X/acd831}
}

@article{togo2023first,
  title={First-principles {P}honon {C}alculations with {P}honopy and {P}hono3py},
  author={Togo, Atsushi},
  journal={J. Phys. Soc. Jpn.},
  volume={92},
  number={1},
  pages={012001},
  year={2023},
  publisher={The Physical Society of Japan},
  doi={https://doi.org/10.7566/JPSJ.92.012001
}
}

@article{Momma:db5098,
author = "Momma, Koichi and Izumi, Fujio",
title = "{{\it VESTA3} for three-dimensional visualization of crystal, volumetric and morphology data}",
journal = "J. Appl. Cryst.",
year = "2011",
volume = "44",
number = "6",
pages = "1272--1276",
month = "Dec",
doi = {10.1107/S0021889811038970},
url = {https://doi.org/10.1107/S0021889811038970},
}

@article{calleja2003trapping,
  title={Trapping of oxygen vacancies on twin walls of \ce{CaTiO3}: a computer simulation study},
  author={Calleja, Mark and Dove, Martin T and Salje, Ekhard KH},
  journal={J. Phys.: Condens. Matter},
  volume={15},
  number={14},
  pages={2301},
  year={2003},
  publisher={IOP Publishing},
  doi={10.1088/0953-8984/15/14/305}
}

@article{barone2014improper,
  title = {Improper origin of polar displacements at \ce{CaTiO3} and \ce{CaMnO3} twin walls},
  author = {Barone, Paolo and Di Sante, Domenico and Picozzi, Silvia},
  journal = {Phys. Rev. B},
  volume = {89},
  issue = {14},
  pages = {144104},
  numpages = {6},
  year = {2014},
  month = {Apr},
  publisher = {American Physical Society},
  doi = {10.1103/PhysRevB.89.144104},
  url = {https://link.aps.org/doi/10.1103/PhysRevB.89.144104}
}

@misc{isotropy,
    author = {Stokes, H. T. Stokes and Hatch, D. M. and Campbell, B. J.},
    title = {ISOTROPY Software Suite},
    howpublished = {\url{iso.byu.edu}}
}

@article{bousquet2016non,
  title={Non-collinear magnetism in multiferroic perovskites},
  author={Bousquet, Eric and Cano, Andr{\'e}s},
  journal={J. Phys.: Condens. Matter },
  volume={28},
  number={12},
  pages={123001},
  year={2016},
  publisher={IOP Publishing},
  doi={10.1088/0953-8984/28/12/123001}
}

@article{bousquet2011induced,
  title = {Induced Magnetoelectric Response in $Pnma$ Perovskites},
  author = {Bousquet, Eric and Spaldin, Nicola},
  journal = {Phys. Rev. Lett.},
  volume = {107},
  issue = {19},
  pages = {197603},
  numpages = {5},
  year = {2011},
  month = {Nov},
  publisher = {American Physical Society},
  doi = {10.1103/PhysRevLett.107.197603},
  url = {https://link.aps.org/doi/10.1103/PhysRevLett.107.197603}
}

@article{benedek2012polar,
  title={Polar octahedral rotations: {A} path to new multifunctional materials},
  author={Benedek, Nicole A and Mulder, Andrew T and Fennie, Craig J},
  journal={J. Solid State Chem.},
  volume={195},
  pages={11--20},
  year={2012},
  publisher={Elsevier},
  doi={https://doi.org/10.1016/j.jssc.2012.04.012}
}

@article{rabe2005theoretical,
  title={Theoretical investigations of epitaxial strain effects in ferroelectric oxide thin films and superlattices},
  author={Rabe, Karin M},
  journal={Curr. Opin. Solid State Mater. Sci.},
  volume={9},
  number={3},
  pages={122--127},
  year={2005},
  publisher={Elsevier},
  doi={https://doi.org/10.1016/j.cossms.2006.06.003}
}

@article{rooj2025altermagnetism,
  title = {Altermagnetism in the orthorhombic $Pnma$ structure through group theory and {DFT} calculations},
  author = {Rooj, Suman and Saxena, Sugandha and Ganguli, Nirmal},
  journal = {Phys. Rev. B},
  volume = {111},
  issue = {1},
  pages = {014434},
  numpages = {11},
  year = {2025},
  month = {Jan},
  publisher = {American Physical Society},
  doi = {10.1103/PhysRevB.111.014434},
  url = {https://link.aps.org/doi/10.1103/PhysRevB.111.014434}
}

@article{KANAMORI195987,
title = {Superexchange interaction and symmetry properties of electron orbitals},
journal = {J. Phys. Chem. Solids},
volume = {10},
number = {2},
pages = {87-98},
year = {1959},
issn = {0022-3697},
doi = {https://doi.org/10.1016/0022-3697(59)90061-7},
url = {https://www.sciencedirect.com/science/article/pii/0022369759900617},
author = {Junjiro Kanamori}
}

@article{Anderson1950,
  title = {Antiferromagnetism. {T}heory of {S}uperexchange {I}nteraction},
  author = {Anderson, P. W.},
  journal = {Phys. Rev.},
  volume = {79},
  issue = {2},
  pages = {350--356},
  numpages = {0},
  year = {1950},
  month = {Jul},
  publisher = {American Physical Society},
  doi = {10.1103/PhysRev.79.350},
  url = {https://link.aps.org/doi/10.1103/PhysRev.79.350}
}

@article{Anderson1959,
  title = {New {A}pproach to the {T}heory of {S}uperexchange {I}nteractions},
  author = {Anderson, P. W.},
  journal = {Phys. Rev.},
  volume = {115},
  issue = {1},
  pages = {2--13},
  numpages = {0},
  year = {1959},
  month = {Jul},
  publisher = {American Physical Society},
  doi = {10.1103/PhysRev.115.2},
  url = {https://link.aps.org/doi/10.1103/PhysRev.115.2}
}

@book{goodenough1963magnetism,
  title={Magnetism and the chemical bond},
  author={Goodenough, John B},
  publisher = { John Wiley \& Sons},
  year={1963}
}

@book{lalena2020principles,
  title={Principles of inorganic materials design},
  author={Lalena, John N and Cleary, David A},
  year={2005},
  publisher={John Wiley \& Sons}
}

@article{Klarbring2018,
  title = {Nature of the octahedral tilting phase transitions in perovskites: A case study of \ce{CaMnO3}},
  author = {Klarbring, Johan and Simak, Sergei I.},
  journal = {Phys. Rev. B},
  volume = {97},
  issue = {2},
  pages = {024108},
  numpages = {10},
  year = {2018},
  month = {Jan},
  publisher = {American Physical Society},
  doi = {10.1103/PhysRevB.97.024108},
  url = {https://link.aps.org/doi/10.1103/PhysRevB.97.024108}
}

@article{Rabe2009,
  title = {Strain-induced ferroelectricity in orthorhombic \ce{CaTiO3} from first principles},
  author = {Eklund, C.-J. and Fennie, C. J. and Rabe, K. M.},
  journal = {Phys. Rev. B},
  volume = {79},
  issue = {22},
  pages = {220101},
  numpages = {4},
  year = {2009},
  month = {Jun},
  publisher = {American Physical Society},
  doi = {10.1103/PhysRevB.79.220101},
  url = {https://link.aps.org/doi/10.1103/PhysRevB.79.220101}
}

@article{aschauer2014competition,
  title={Competition and cooperation between antiferrodistortive and ferroelectric instabilities in the model perovskite \ce{SrTiO3}},
  author={Aschauer, Ulrich and Spaldin, Nicola A},
  journal={J. Phys.: Condens. Matter},
  volume={26},
  number={12},
  pages={122203},
  year={2014},
  publisher={IOP Publishing},
  doi={10.1088/0953-8984/26/12/122203}
}

@article{Terakura2010,
  title = {First-principles calculations for the magnetic phase diagram in electron-doped \ce{CaMnO3} under compressive and tensile strains},
  author = {Tsukahara, Hiroshi and Ishibashi, Shoji and Terakura, Kiyoyuki},
  journal = {Phys. Rev. B},
  volume = {81},
  issue = {21},
  pages = {214108},
  numpages = {8},
  year = {2010},
  month = {Jun},
  publisher = {American Physical Society},
  doi = {10.1103/PhysRevB.81.214108},
  url = {https://link.aps.org/doi/10.1103/PhysRevB.81.214108}
}

@article{Marthinsen2016, 
title={Coupling and competition between ferroelectricity, magnetism, strain, and oxygen vacancies in \ce{AMnO3} perovskites}, 
volume={6}, 
DOI={10.1557/mrc.2016.30}, 
number={3}, 
journal={MRS Commun.}, 
author={Marthinsen, Astrid and Faber, Carina and Aschauer, Ulrich and Spaldin, Nicola A. and Selbach, Sverre M.}, 
year={2016}, 
pages={182–191}}

@article{Vanderbilt2012,
  title = {Spin-phonon coupling effects in transition-metal perovskites: A {DFT} + ${U}$ and hybrid-functional study},
  author = {Hong, Jiawang and Stroppa, Alessandro and \'I\~niguez, Jorge and Picozzi, Silvia and Vanderbilt, David},
  journal = {Phys. Rev. B},
  volume = {85},
  issue = {5},
  pages = {054417},
  numpages = {12},
  year = {2012},
  month = {Feb},
  publisher = {American Physical Society},
  doi = {10.1103/PhysRevB.85.054417},
  url = {https://link.aps.org/doi/10.1103/PhysRevB.85.054417}
}

@article{Andersen1991,
  title = {Band theory and {M}ott insulators: {H}ubbard ${U}$ instead of {S}toner ${I}$},
  author = {Anisimov, Vladimir I. and Zaanen, Jan and Andersen, Ole K.},
  journal = {Phys. Rev. B},
  volume = {44},
  issue = {3},
  pages = {943--954},
  numpages = {0},
  year = {1991},
  month = {Jul},
  publisher = {American Physical Society},
  doi = {10.1103/PhysRevB.44.943},
  url = {https://link.aps.org/doi/10.1103/PhysRevB.44.943}
}

@article{zhao2019creating,
  title={Creating multiferroic and conductive domain walls in common ferroelastic compounds},
  author={Zhao, Hong Jian and {\'I}{\~n}iguez, Jorge},
  journal={npj Comput. Mater.},
  volume={5},
  number={1},
  pages={92},
  year={2019},
  publisher={Nature Publishing Group UK London},
  doi={https://doi.org/10.1038/s41524-019-0229-5}
}

@book{salje1990phase,
  title={Phase transitions in ferroelastic and co-elastic crystals},
  author={Salje, Ekhard},
  year={1990},
  publisher={Cambridge University Press}
}

@article{khalyavin2015antisymmetric,
  title={Antisymmetric exchange in \ce{La}-substituted \ce{BiFe_{0.5}Sc_{0.5}O3} system: symmetry adapted distortion modes approach},
  author={Khalyavin, Dmitry D and Salak, Andrei N and Manuel, Pascal and Olekhnovich, Nikolai M and Pushkarev, Anatoly V and Radysh, Yury V and Fedorchenko, Alexey V and Fertman, Elena L and Desnenko, Vladimir A and Ferreira, M{\'a}rio GS},
  journal={Z. Kristallogr. Cryst. Mater.},
  volume={230},
  number={12},
  pages={767--774},
  year={2015},
  publisher={De Gruyter},
  doi={https://doi.org/10.1515/zkri-2015-1873}
}

@article{senn2018group,
  title={A group-theoretical approach to enumerating magnetoelectric and multiferroic couplings in perovskites},
  author={Senn, Mark S and Bristowe, Nicholas C},
  journal={Acta Crystallogr. Sect. A},
  volume={74},
  number={4},
  pages={308--321},
  year={2018},
  publisher={International Union of Crystallography},
  doi={https://doi.org/10.1107/S2053273318007441}
}

@article{Weingart2012,
  title = {Noncollinear magnetism and single-ion anisotropy in multiferroic perovskites},
  author = {Weingart, Carlo and Spaldin, Nicola and Bousquet, Eric},
  journal = {Phys. Rev. B},
  volume = {86},
  issue = {9},
  pages = {094413},
  numpages = {11},
  year = {2012},
  month = {Sep},
  publisher = {American Physical Society},
  doi = {10.1103/PhysRevB.86.094413},
  url = {https://link.aps.org/doi/10.1103/PhysRevB.86.094413}
}

@article{Zhou2020,
  title = {Weak ferromagnetism in perovskite oxides},
  author = {Zhou, J.-S. and Marshall, L. G. and Li, Z.-Y. and Li, X. and He, J.-M.},
  journal = {Phys. Rev. B},
  volume = {102},
  issue = {10},
  pages = {104420},
  numpages = {6},
  year = {2020},
  month = {Sep},
  publisher = {American Physical Society},
  doi = {10.1103/PhysRevB.102.104420},
  url = {https://link.aps.org/doi/10.1103/PhysRevB.102.104420}
}

@article{zhou2013strain,
  title={Strain-induced hybrid improper ferroelectricity in simple perovskites from first principles},
  author={Zhou, Qibin and Rabe, Karin M},
  journal={arXiv preprint arXiv:1306.1839},
  year={2013},
  url={https://doi.org/10.48550/arXiv.1306.1839}
}

@article{Dagotto2006,
  title = {Role of the {D}zyaloshinskii-{M}oriya interaction in multiferroic perovskites},
  author = {Sergienko, I. A. and Dagotto, E.},
  journal = {Phys. Rev. B},
  volume = {73},
  issue = {9},
  pages = {094434},
  numpages = {5},
  year = {2006},
  month = {Mar},
  publisher = {American Physical Society},
  doi = {10.1103/PhysRevB.73.094434},
  url = {https://link.aps.org/doi/10.1103/PhysRevB.73.094434}
}

@article{Kalinin2012,
  title = {Conductivity of twin-domain-wall/surface junctions in ferroelastics: Interplay of deformation potential, octahedral rotations, improper ferroelectricity, and flexoelectric coupling},
  author = {Eliseev, Eugene A. and Morozovska, Anna N. and Gu, Yijia and Borisevich, Albina Y. and Chen, Long-Qing and Gopalan, Venkatraman and Kalinin, Sergei V.},
  journal = {Phys. Rev. B},
  volume = {86},
  issue = {8},
  pages = {085416},
  numpages = {10},
  year = {2012},
  month = {Aug},
  publisher = {American Physical Society},
  doi = {10.1103/PhysRevB.86.085416},
  url = {https://link.aps.org/doi/10.1103/PhysRevB.86.085416}
}

@article{Gopalan2012,
  title = {Interfacial polarization and pyroelectricity in antiferrodistortive structures induced by a flexoelectric effect and rotostriction},
  author = {Morozovska, Anna N. and Eliseev, Eugene A. and Glinchuk, Maya D. and Chen, Long-Qing and Gopalan, Venkatraman},
  journal = {Phys. Rev. B},
  volume = {85},
  issue = {9},
  pages = {094107},
  numpages = {9},
  year = {2012},
  month = {Mar},
  publisher = {American Physical Society},
  doi = {10.1103/PhysRevB.85.094107},
  url = {https://link.aps.org/doi/10.1103/PhysRevB.85.094107}
}

@article{Stengel2017,
  title = {Macroscopic {P}olarization from {A}ntiferrodistortive {C}ycloids in {F}erroelastic \ce{SrTiO3}},
  author = {Schiaffino, Andrea and Stengel, Massimiliano},
  journal = {Phys. Rev. Lett.},
  volume = {119},
  issue = {13},
  pages = {137601},
  numpages = {5},
  year = {2017},
  month = {Sep},
  publisher = {American Physical Society},
  doi = {10.1103/PhysRevLett.119.137601},
  url = {https://link.aps.org/doi/10.1103/PhysRevLett.119.137601}
}

@article{Fennie2011,
  title = {Hybrid {I}mproper {F}erroelectricity: {A} {M}echanism for {C}ontrollable {P}olarization-{M}agnetization {C}oupling},
  author = {Benedek, Nicole A. and Fennie, Craig J.},
  journal = {Phys. Rev. Lett.},
  volume = {106},
  issue = {10},
  pages = {107204},
  numpages = {4},
  year = {2011},
  month = {Mar},
  publisher = {American Physical Society},
  doi = {10.1103/PhysRevLett.106.107204},
  url = {https://link.aps.org/doi/10.1103/PhysRevLett.106.107204}
}

@article{Dey2025,
  title = {Prediction of {R}oom {T}emperature {E}lectric {F}ield {R}eversal of {M}agnetization in the {F}amily of \ce{A4B3O9} {L}ayered {O}xides},
  author = {Dey, Urmimala and McCabe, Emma E. and \'I\~niguez-Gonz\'alez, Jorge and Bristowe, Nicholas C.},
  journal = {Phys. Rev. Lett.},
  volume = {134},
  issue = {13},
  pages = {136801},
  numpages = {7},
  year = {2025},
  month = {Apr},
  publisher = {American Physical Society},
  doi = {10.1103/PhysRevLett.134.136801},
  url = {https://link.aps.org/doi/10.1103/PhysRevLett.134.136801}
}

@article{aoyama2014giant,
  title={Giant spin-driven ferroelectric polarization in \ce{TbMnO3} under high pressure},
  author={Aoyama, T and Yamauchi, K and Iyama, A and Picozzi, S and Shimizu, K and Kimura, T},
  journal={Nat. Commun.},
  volume={5},
  number={1},
  pages={4927},
  year={2014},
  publisher={Nature Publishing Group UK London},
  doi={https://doi.org/10.1038/ncomms5927}
}

@article{dagotto2008strongly,
  title={Strongly {C}orrelated {E}lectronic {M}aterials: {P}resent and {F}uture},
  author={Dagotto, E and Tokura, Y},
  journal={MRS Bull.},
  volume={33},
  number={11},
  pages={1037--1045},
  year={2008},
  publisher={Cambridge University Press},
 doi={https://doi.org/10.1557/mrs2008.223}
}

@article{terakura2007magnetism,
  title={Magnetism, orbital ordering and lattice distortion in perovskite transition-metal oxides},
  author={Terakura, K},
  journal={Prog. Mater. Sci.},
  volume={52},
  number={2-3},
  pages={388--400},
  year={2007},
  publisher={Elsevier},
  doi={https://doi.org/10.1016/j.pmatsci.2006.10.007}
}

@article{goodenough2014perspective,
  title={Perspective on {E}ngineering {T}ransition-{T}etal {O}xides},
  author={Goodenough, John B},
  journal={Chem. Mater.},
  volume={26},
  number={1},
  pages={820--829},
  year={2014},
  publisher={ACS Publications}
}

@article{aschauer2016interplay,
  title={Interplay between strain, defect charge state, and functionality in complex oxides},
  author={Aschauer, Ulrich and Spaldin, Nicola A},
  journal={Appl. Phys. Lett.},
  volume={109},
  number={3},
  year={2016},
  publisher={AIP Publishing},
  url={https://doi.org/10.1063/1.4958716}
}

@article{tokura2010multiferroics,
  title={Multiferroics with {S}piral {S}pin {O}rders},
  author={Tokura, Yoshinori and Seki, Shinichiro},
  journal={Adv. Mater.},
  volume={22},
  number={14},
  pages={1554--1565},
  year={2010},
  publisher={Wiley Online Library},
  doi={https://doi.org/10.1002/adma.200901961}
}

@article{Xiang2013,
  title = {Unified model of ferroelectricity induced by spin order},
  author = {Xiang, H. J. and Wang, P. S. and Whangbo, M.-H. and Gong, X. G.},
  journal = {Phys. Rev. B},
  volume = {88},
  issue = {5},
  pages = {054404},
  numpages = {6},
  year = {2013},
  month = {Aug},
  publisher = {American Physical Society},
  doi = {10.1103/PhysRevB.88.054404},
  url = {https://link.aps.org/doi/10.1103/PhysRevB.88.054404}
}

@article{Moriya1960,
  title = {Anisotropic {S}uperexchange {I}nteraction and {W}eak {F}erromagnetism},
  author = {Moriya, T\^oru},
  journal = {Phys. Rev.},
  volume = {120},
  issue = {1},
  pages = {91--98},
  numpages = {0},
  year = {1960},
  month = {Oct},
  publisher = {American Physical Society},
  doi = {10.1103/PhysRev.120.91},
  url = {https://link.aps.org/doi/10.1103/PhysRev.120.91}
}

@article{dzyaloshinsky1958thermodynamic,
  title={A thermodynamic theory of “weak” ferromagnetism of antiferromagnetics},
  author={Dzyaloshinsky, Igor},
  journal={J. Phys. Chem. Solids},
  volume={4},
  number={4},
  pages={241--255},
  year={1958},
  publisher={Elsevier},
  doi={https://doi.org/10.1016/0022-3697(58)90076-3}
}

@article{meier2022ferroelectric,
  title={Ferroelectric domain walls for nanotechnology},
  author={Meier, Dennis and Selbach, Sverre M},
  journal={Nat. Rev. Mater.},
  volume={7},
  number={3},
  pages={157--173},
  year={2022},
  publisher={Nature Publishing Group UK London},
  doi={https://doi.org/10.1038/s41578-021-00375-z}
}

@article{zhou2015ferroelectricity,
  title={Ferroelectricity driven magnetism at domain walls in \ce{LaAlO3}/\ce{PbTiO3} superlattices},
  author={Zhou, PX and Dong, S and Liu, HM and Ma, CY and Yan, ZB and Zhong, CG and Liu, J-M},
  journal={Sci. Rep.},
  volume={5},
  number={1},
  pages={13052},
  year={2015},
  publisher={Nature Publishing Group UK London},
  doi={https://doi.org/10.1038/srep13052}
}

@article{geng2012collective,
  title={Collective {M}agnetism at {M}ultiferroic {V}ortex {D}omain {W}alls},
  author={Geng, Yanan and Lee, Nara and Choi, YJ and Cheong, S-W and Wu, Weida},
  journal={Nano Lett.},
  volume={12},
  number={12},
  pages={6055--6059},
  year={2012},
  publisher={ACS Publications},
  doi={https://doi.org/10.1021/nl301432z}
}

@article{Daraktchiev2010,
  title = {Landau theory of domain wall magnetoelectricity},
  author = {Daraktchiev, Maren and Catalan, Gustau and Scott, James F.},
  journal = {Phys. Rev. B},
  volume = {81},
  issue = {22},
  pages = {224118},
  numpages = {12},
  year = {2010},
  month = {Jun},
  publisher = {American Physical Society},
  doi = {10.1103/PhysRevB.81.224118},
  url = {https://link.aps.org/doi/10.1103/PhysRevB.81.224118}
}

@article{geirhos2020macroscopic,
  title={Macroscopic manifestation of domain-wall magnetism and magnetoelectric effect in a {N}{\'e}el-type skyrmion host},
  author={Geirhos, Korbinian and Gross, Boris and Szigeti, Bertalan G and Mehlin, Andrea and Philipp, Simon and White, Jonathan S and Cubitt, Robert and Widmann, Sebastian and Ghara, Somnath and Lunkenheimer, Peter and others},
  journal={npj Quantum Mater.},
  volume={5},
  number={1},
  pages={44},
  year={2020},
  publisher={Nature Publishing Group UK London},
  doi={https://doi.org/10.1038/s41535-020-0247-z}
}

@article{farokhipoor2014artificial,
  title={Artificial chemical and magnetic structure at the domain walls of an epitaxial oxide},
  author={Farokhipoor, S and Mag{\'e}n, C{\'e}sar and Venkatesan, Sriram and {\'I}{\~n}iguez, Jorge and Daumont, Christophe JM and Rubi, Diego and Snoeck, Etienne and Mostovoy, M and De Graaf, Coen and M{\"u}ller, Alexander and others},
  journal={Nature},
  volume={515},
  number={7527},
  pages={379--383},
  year={2014},
  publisher={Nature Publishing Group UK London},
  doi={https://doi.org/10.1038/nature13918}
}

@article{cheong2020seeing,
  title={Seeing is believing: Visualization of antiferromagnetic domains},
  author={Cheong, Sang-Wook and Fiebig, Manfred and Wu, Weida and Chapon, Laurent and Kiryukhin, Valery},
  journal={npj Quantum Mater.},
  volume={5},
  number={1},
  pages={3},
  year={2020},
  publisher={Nature Publishing Group UK London},
  doi={https://doi.org/10.1038/s41535-019-0204-x}
}

@article{geng2014direct,
  title={Direct visualization of magnetoelectric domains},
  author={Geng, Yanan and Das, Hena and Wysocki, Aleksander L and Wang, Xueyun and Cheong, SW and Mostovoy, M and Fennie, Craig J and Wu, Weida},
  journal={Nat. Mater.},
  volume={13},
  number={2},
  pages={163--167},
  year={2014},
  publisher={Nature Publishing Group UK London},
  doi={https://doi.org/10.1038/nmat3813}
}

@article{zhang2023first,
  title={First-principles calculations of domain wall energies of prototypical ferroelectric perovskites},
  author={Zhang, Xueyou and Wang, Bo and Ji, Yanzhou and Xue, Fei and Wang, Yi and Chen, Long-Qing and Nan, Ce-Wen},
  journal={Acta Mater.},
  volume={242},
  pages={118351},
  year={2023},
  publisher={Elsevier},
  doi={https://doi.org/10.1016/j.actamat.2022.118351}
}

@article{eggestad2024mobile,
  title={Mobile intrinsic point defects for conductive neutral domain walls in \ce{LiNbO3}},
  author={Eggestad, Kristoffer and Williamson, Benjamin A D and Meier, Dennis and Selbach, Sverre M},
  journal={J. Mater. Chem. C},
  volume={12},
  number={42},
  pages={17099--17107},
  year={2024},
  publisher={Royal Society of Chemistry},
  doi={10.1039/D4TC02856B}
}

@article{wei2014ferroelectric,
  title={Ferroelectric translational antiphase boundaries in nonpolar materials},
  author={Wei, Xian-Kui and Tagantsev, Alexander K and Kvasov, Alexander and Roleder, Krystian and Jia, Chun-Lin and Setter, Nava},
  journal={Nat. Commun.},
  volume={5},
  number={1},
  pages={3031},
  year={2014},
  publisher={Nature Publishing Group UK London},
  doi={https://doi.org/10.1038/ncomms4031}
}

@article{chauleau2020electric,
  title={Electric and antiferromagnetic chiral textures at multiferroic domain walls},
  author={Chauleau, J-Y and Chirac, Th{\'e}ophile and Fusil, St{\'e}phane and Garcia, Vincent and Akhtar, W and Tranchida, Julien and Thibaudeau, Pascal and Gross, Isabell and Blouzon, Camille and Finco, Aurore and others},
  journal={Nat. Mater.},
  volume={19},
  number={4},
  pages={386--390},
  year={2020},
  publisher={Nature Publishing Group UK London},
  doi={https://doi.org/10.1038/s41563-019-0516-z}
}

@article{han2019topological,
  title={Topological {M}agnetic-{S}pin {T}extures in {T}wo-{D}imensional van der {W}aals \ce{Cr2Ge2Te6}},
  author={Han, Myung-Geun and Garlow, Joseph A and Liu, Yu and Zhang, Huiqin and Li, Jun and DiMarzio, Donald and Knight, Mark W and Petrovic, Cedomir and Jariwala, Deep and Zhu, Yimei},
  journal={Nano Lett.},
  volume={19},
  number={11},
  pages={7859--7865},
  year={2019},
  publisher={ACS Publications},
  doi={https://doi.org/10.1021/acs.nanolett.9b02849}
}

@article{lu2025temperature,
  title={Temperature dependence of emerging properties of ferroelastic domain walls in \ce{CaTiO3}},
  author={Lu, Guangming and Salje, Ekhard KH},
  journal={J. Appl. Phys.},
  volume={137},
  number={15},
  year={2025},
  publisher={AIP Publishing},
  url={https://doi.org/10.1063/5.0258013}
}

@article{zemp2024magnetoelectric,
  title = {Magnetoelectric coupling in the multiferroic hybrid-improper ferroelectric \ce{Ca3Mn_{1.9}Ti_{0.1}O7}},
  author = {Zemp, Yannik and Trassin, Morgan and Gradauskaite, Elzbieta and Gao, Bin and Cheong, Sang-Wook and Lottermoser, Thomas and Fiebig, Manfred and Weber, Mads C.},
  journal = {Phys. Rev. B},
  volume = {109},
  issue = {18},
  pages = {184417},
  numpages = {8},
  year = {2024},
  month = {May},
  publisher = {American Physical Society},
  doi = {10.1103/PhysRevB.109.184417},
  url = {https://link.aps.org/doi/10.1103/PhysRevB.109.184417}
}

\end{document}

\preprint{AIP/123-QED}

\title{Supplemental material: \\ Local indirect magnetoelectric coupling at twin walls in \ce{CaMnO3}}

\author{Ida C. Skogvoll}
 \affiliation{Department of Materials Science and Engineering, Faculty of Natural Sciences and Technology,
    NTNU Norwegian University of Science and Technology, NO-7491 Trondheim, Norway}%Lines break automatically or can be forced with \\
\author{Benjamin A. D. Williamson}
\affiliation{%
Department of Materials Science and Engineering, Faculty of Natural Sciences and Technology,
    NTNU Norwegian University of Science and Technology, NO-7491 Trondheim, Norway}%\\This line break forced% with \\
\author{Sverre M. Selbach}%
 \email{selbach@ntnu.no}
\affiliation{Department of Materials Science and Engineering, Faculty of Natural Sciences and Technology,
    NTNU Norwegian University of Science and Technology, NO-7491 Trondheim, Norway}%

{%
%\\This line break forced% with \\
}%

\date{\today}% It is always \today, today,

\maketitle

\section{Phonon calculations}
\begin{figure}[h]
    \centering
    \includegraphics[width=1\linewidth]{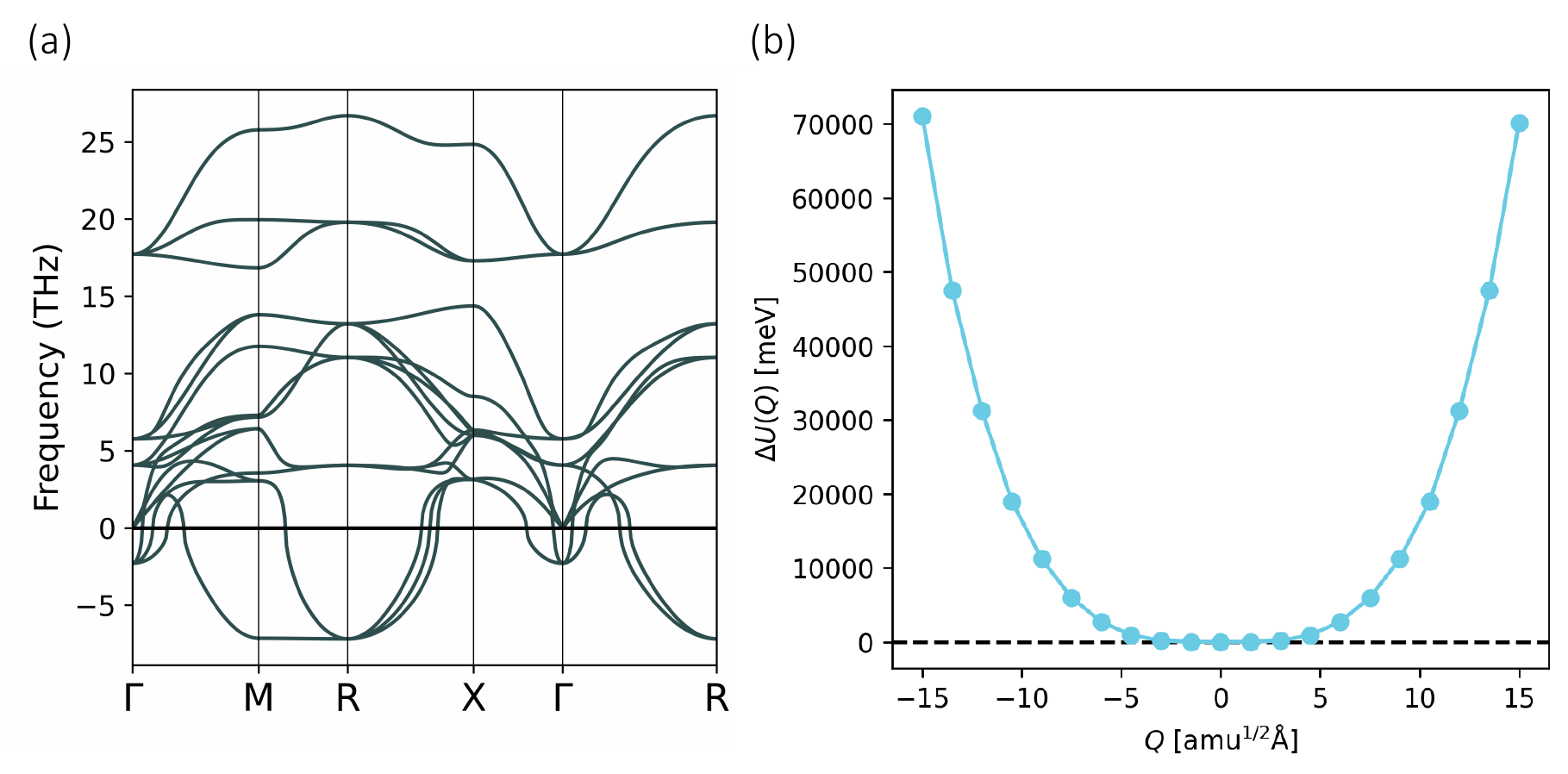}
    \caption{(a) Calculated phonon spectrum for the cubic $Pm\bar{3}m$ phase of \ce{CaMnO3}.  (b) Relative energy difference as a function of the distortion mode amplitude of the $\Gamma_4^-$. Although the base of the well is wide, there is only a single minimum at the center.}
    \label{fig:SM1}
\end{figure}

The phonon spectrum for \ce{CaMnO3} was calculated, resulting in imaginary phonons at the M and R-mode of the Brillouin zone, corresponding to the in- and out-of-phase rotations of the \ce{MnO6}-octahedra. In addition, there is a smaller instability in the $\Gamma$-point which contains a B-site displacement away from the octahedral center-of-mass. Calculating the energy as a function of this $\Gamma$-mode displacement yields no energy lowering, consistent with a B-site instability which is suppressed in the bulk, see Figure \ref{fig:SM1}. 

\newpage

\section{Collinear magnetic domain walls}
\subsection{Formation energies}
\begin{figure}[h]
    \centering
    \includegraphics[width=1\linewidth]{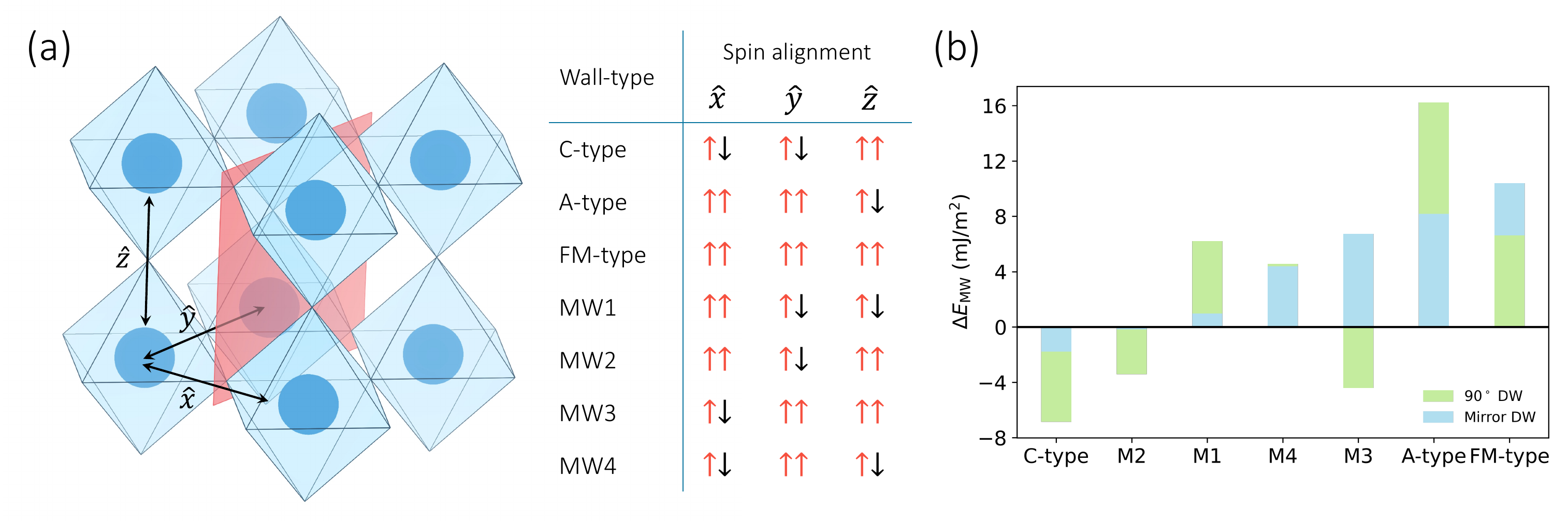}
    \caption{(a) The \ce{Mn} atoms are designated by blue spheres and the center of the ferroelastic DW indicated by the red plane. The adjoined table lists the different MW configurations with spin orders along each direction. (c) The calculated energy difference associated with putting a MW at the DW as opposed to the bulk. Note that for MW2, the formation energy associated with the mirror DW is approximately 0, while for MW4 the different energies are close to equal.}
    \label{fig:SM2}
\end{figure}

The susceptibility of the different domain wall (DW) types towards a localized magnetic order is tested by calculating the formation energies using Eq. 5 in the main text. This local magnetic ordering is referred to as a magnetic wall (MW), with the eight possible spin configurations displayed in Figure \ref{fig:SM2}(a). Each MW is restricted to the two layers of \ce{Mn} atoms adjacent and parallel to the center of the structural DW. The resulting energies are listed in Figure \ref{fig:SM2}(b). Following the rules for \ce{d^3-d^3} superexchange as outlined by Goodenough and Kanamori\cite{KANAMORI195987, goodenough1963magnetism}, the exchange coupling is expected to get more ferromagnetic for bonding angles approaching $90^\circ$ and shorter bond lengths, as this leads to a smaller orbital overlap. This trend is also reflected in the formation energy for each of the MW configurations. 

For both the $90^\circ$ DW and the mirror DW, the C-type wall produces the largest energy-lowering relative to the bulk. This is a natural consequence for the $90^\circ$ DW, as the \ce{Mn-O-Mn} angles become narrower and the \ce{Mn-O} bond lengths become longer in the $z$ direction. For the mirror DW, this is the only significant negative formation energy, and also almost four times lower in absolute value than the $90^\circ$ DW. This can be attributed to the fact that the local DW geometry of this wall does not show any lengthening of the \ce{Mn-O} bonds in any direction, and all \ce{Mn-O-Mn} angles become more straight. In addition, the geometry at the $90^\circ$ is such that both $y$ and $x$ directions should feature a high energy cost for parallel spins, while the $y$ direction is the most unfavorable for the mirror DW. This, along with the more significant energy gain of the latter wall with one ferromagnetic node in the $z$ direction, provides an overall account for these results. 

\subsection{B-site displacements}
\begin{figure}[h]
    \centering
    \includegraphics[width=0.9\linewidth]{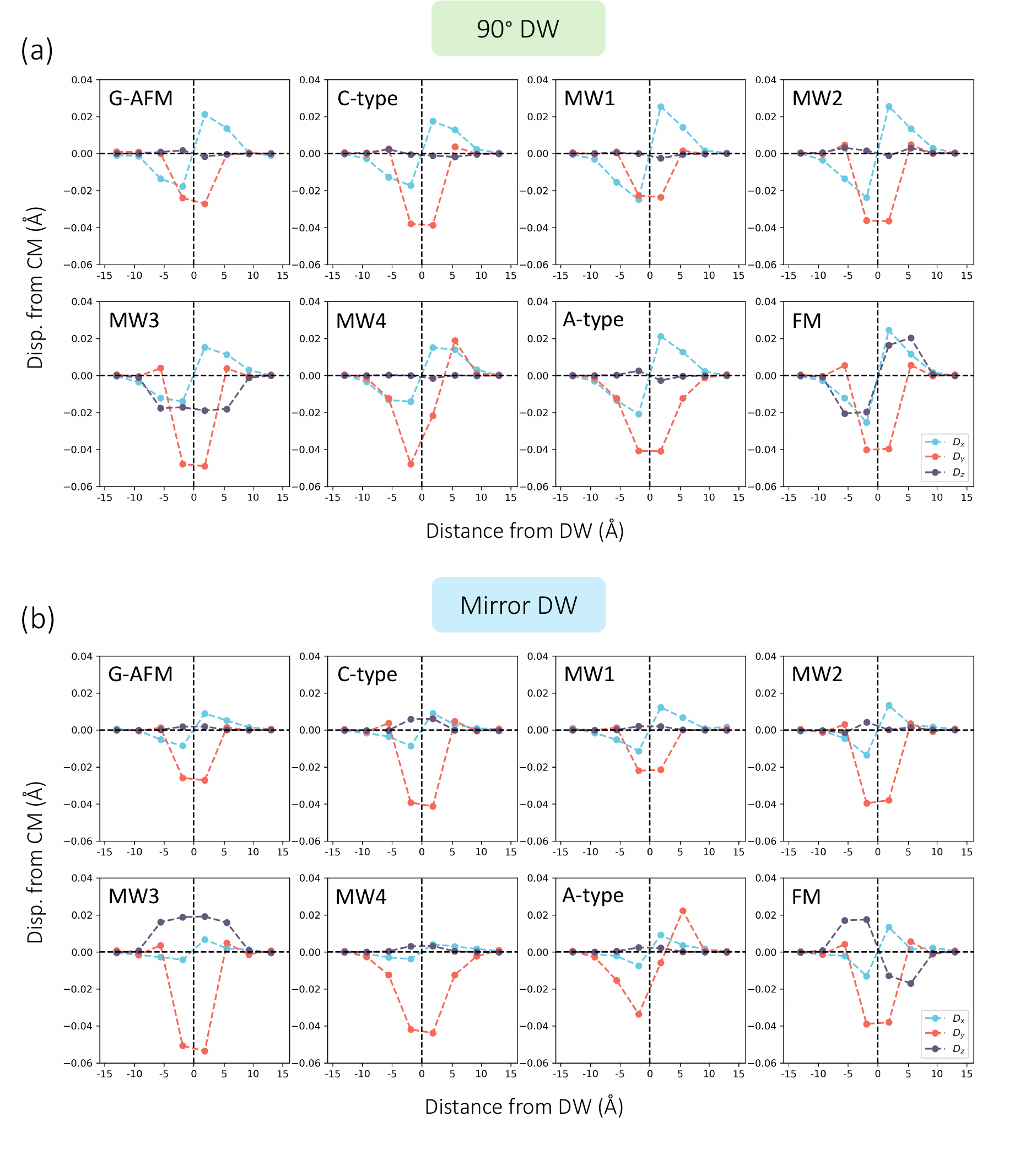}
    \caption{Average displacement of the \ce{Mn} atoms from the octahedron center of mass, as a function of the distance from the first DW, for the (a) $90^\circ$ DW and the (b) mirror DW. The second DW (in the same supercell) shows the same displacement only with $D_y$ switched for the former and $D_x$ and $D_y$ switched for the latter DW type.}
    \label{fig:SM3}
\end{figure}

Changing the magnetic order at the wall directly affects the polar distortion profile, as can be clearly seen in Figure \ref{fig:SM3}. Depending on whether a particular \ce{Mn-O-Mn} bond involves parallel or antiparallel spins, the bond lengths and angles will be altered through the superexchange interaction. Because of the rigid corner-sharing octahedra, the $x$, $y$ and $z$-decomposed polar distortion does not correlate directly with the particular spin ordering along that corresponding axis. The magnitude of the induced polar distortion also depends on the rotational mode being modified, which again ties to the particular symmetry-breaking of each of the $R$ and $M$-modes at each wall. The various MWs therefore give quantitatively and qualitatively different polarization profiles, which also differ between the twin wall types. 

In common for both domain wall types is that all MWs with parallel spins along the $z$-axis, C-type, FM-type, MW2 and MW3, give an enhanced polar distortion of the \ce{Mn}-atoms in the $y$ direction ($D_y$). In addition, the MWs with $x$ direction parallel spins also see a $D_x$ enhancement. MW3 and FM-type, which differ by the ordering of the $x$ direction spins, see the same anomalous $D_z$ distortion, only with a switching at the boundary for FM. It is challenging to detangle exactly how each \ce{Mn}-distortion couples to the different spin orders through these calculations, but the consistent enhancement of the polar distortion with parallel spins in the out-of-plane orientation suggests that the displacement is most strongly coupled to the out-of-phase octahedral rotations. These are the ones mostly affected by an alteration of the out-of-plane \ce{Mn-O-Mn} bonds. 

\newpage
\section{Non-collinear magnetic calculations}

\begin{figure}[h]
    \centering
    \includegraphics[width=0.8\linewidth]{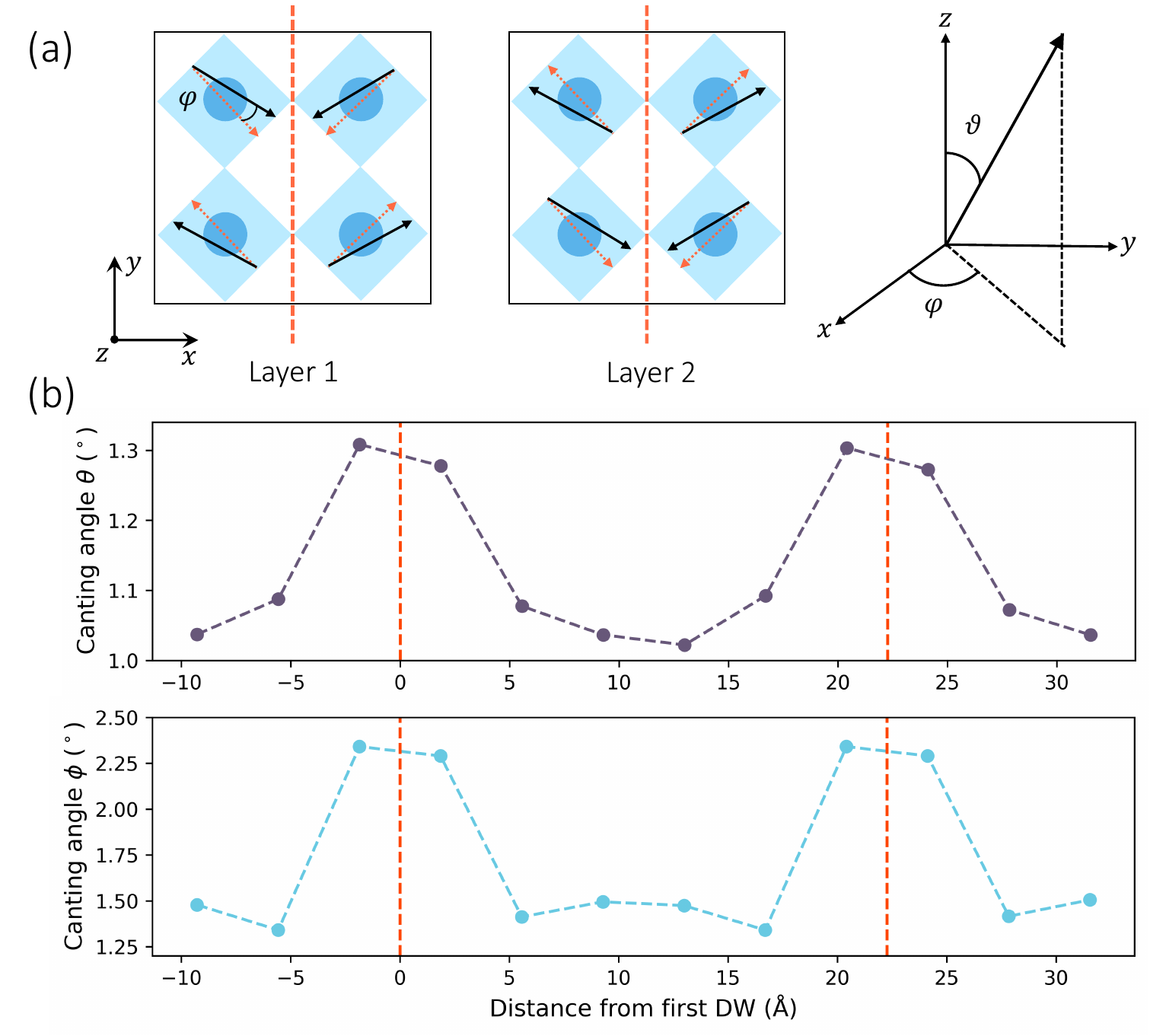}
    \caption{(a) Black arrows show the canting of each spin away from the uncanted initial state (red dashed arrow). Here the interface corresponds to a mirror operation. Vertical red dashed lines indicate the location of the DW, for each of the two layers of octahedra in the $xy$-plane of the supercell. (b) The average canting angle of the spins as a function of distance from the DW, defined in terms of spherical coordinates $\theta$ and $\phi$. }
    \label{fig:SM4}
\end{figure}

\begin{figure}[h]
    \centering
    \includegraphics[width=0.8\linewidth]{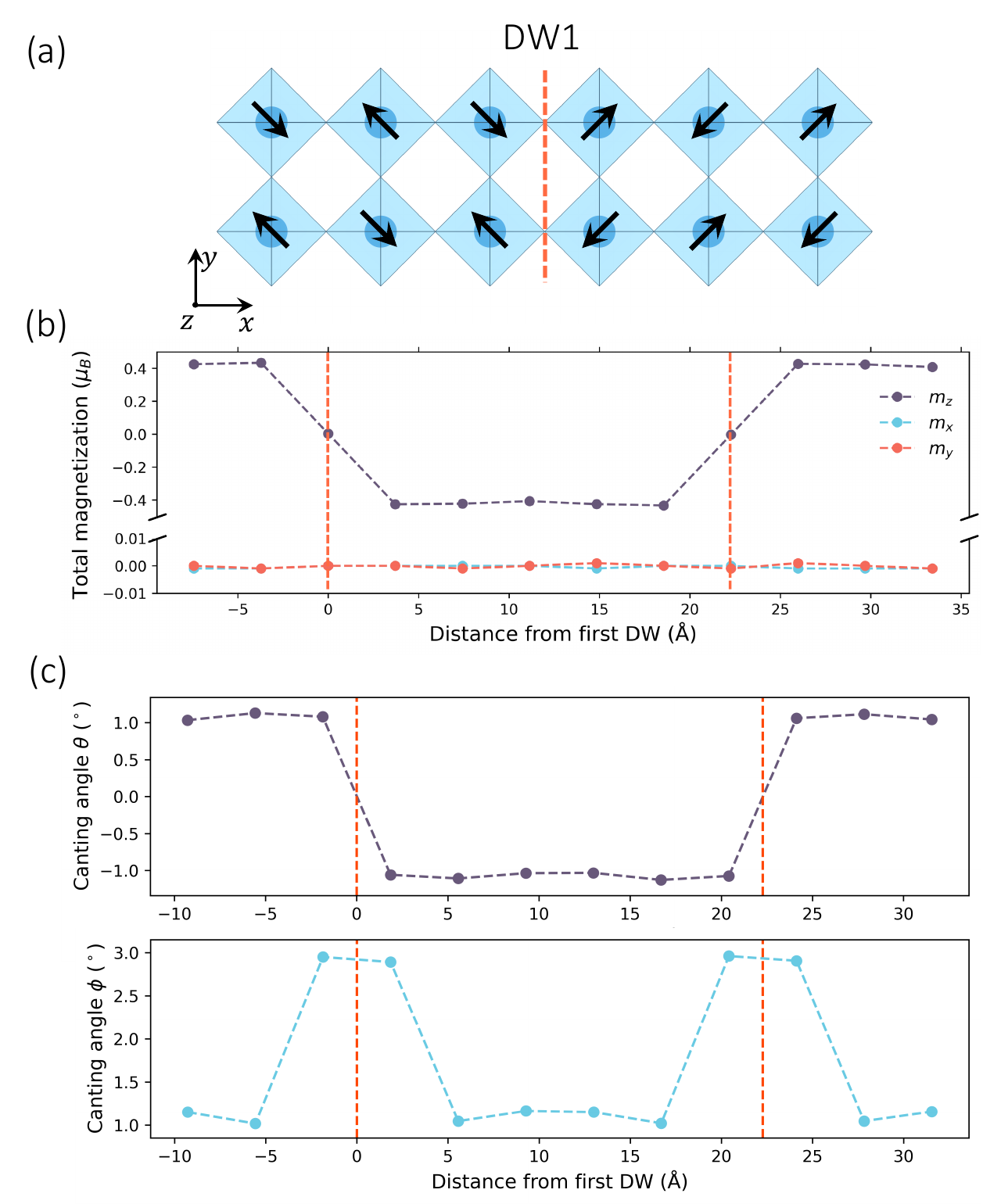}
    \caption{(a) The initial magnetic order prior to the self-consistent spin rotation. Here the interface corresponds to a translation along with a $90^\circ$ rotation (b) Average magnetization of one unit cell consisting of eight \ce{Mn} atoms. Note the break in the y-axis. (c) The average canting angle of the spins as a function of distance from the DW, defined in terms of spherical coordinates $\theta$ and $\phi$.  }
    \label{fig:SM5}
\end{figure}

For the non-collinear calculations, a magnetic order was chosen where the direction determined by the magnetic anisotropy energy is conserved within each domain. We assume this to be the most correct approach, despite working with a smaller $6\times 1\times 1$ supercell. This results in two alternatives for the interface, one where there is mirror symmetry and one corresponding to a translation along with a $90^\circ$ rotation of each spin. The latter was determined to be of lower energy with a difference of \SI{0.01}{meV/f.u.}. The canting angles for each alternative are shown in Figs. \ref{fig:SM4} and \ref{fig:SM5}, together with the magnetic configuration and the average magnetization for the higher energy alternative in Figure \ref{fig:SM5}. 

When allowing the spins to rotate self-consistently, both configurations resulted in a larger canting at the DW, but with the rotation-symmetric magnetic interface resulting in a reversal of the $m_z$ component between each domain. This is because the two different interfaces are connected by a time-reversal of the mid-domain spin order. 

\newpage
\bibliography{apssamp}% Produces the bibliography via BibTeX.